\documentclass[twocolumn,showpacs,preprintnumbers,amsmath,amssymb,superscriptaddress,balancelastpage,prb]{revtex4}
\include{options}
\usepackage {amsmath,amssymb,epsfig,color}
\usepackage{natbib}
\usepackage {graphicx}

\begin{document}

\title{Polaron transformations in the realistic model of the strongly correlated electron system}
\author{E.~I.~Shneyder}
\email{eshneyder@gmail.com}
\affiliation{Kirensky Institute of Physics, Federal Research Center KSC SB RAS, 660036 Krasnoyarsk, Russia}
\affiliation{Reshetnev Siberian State University of Science and Technology, 660037 Krasnoyarsk, Russia}
\author{S.~V.~Nikolaev}
\affiliation{Kirensky Institute of Physics, Federal Research Center KSC SB RAS, 660036 Krasnoyarsk, Russia}
\affiliation{Siberian Federal University, 660041 Krasnoyarsk, Russia}
\author{M.~V.~Zotova}
\affiliation{Kirensky Institute of Physics, Federal Research Center KSC SB RAS, 660036 Krasnoyarsk, Russia}
\affiliation{Siberian Federal University, 660041 Krasnoyarsk, Russia}
\author{R.~A.~Kaldin}
\affiliation{Siberian Federal University, 660041 Krasnoyarsk, Russia}
\author{S.~G.~Ovchinnikov}
\affiliation{Kirensky Institute of Physics, Federal Research Center KSC SB RAS, 660036 Krasnoyarsk, Russia}
\affiliation{Siberian Federal University, 660041 Krasnoyarsk, Russia}
\date{\today}

\begin {abstract}
Electron-phonon coupling, diagonal in a real space formulation, leads to polaron paradigm of smoothly varying properties. However, fundamental changes, namely the singular behavior of polarons, occur if off-diagonal pairing is involved into consideration. The study of polaron transformations and related properties of matter is of particular interest for realistic models, since competition between diagonal and off-diagonal electron-phonon contributions in the presence of other strong interactions can result in unconventional behavior of the system. Here we consider two-dimensional multiband pd-model of cuprate superconductors  with electron-phonon interaction and analyze the features of the systems that are caused by the competition of diagonal and off-diagonal electron-phonon contributions in the limit of strong electron correlations. Using the polaronic version of the generalized tight-binding method, we describe the evolution of the band structure, Fermi surface, density of states at Fermi level, and phonon spectral function in the space of electron-phonon parameters ranging from weak to strong coupling strength of the adiabatic limit. On the phase diagram of polaron properties we reveal two quantum phase transitions and show how electron-phonon interaction gives rise to Fermi surface transformation (i) from hole pockets to true Fermi arcs and (ii) from hole to electron type of conductivity. We also demonstrate the emergence of new states in the phonon spectral function of the polaron and discuss their origin.
\end {abstract}

\maketitle

\section{Introduction \label{intro}}
The concept of polaron has been of broad interest in physics since the appearance of the fundamental papers of Landau and Pekar \cite{Landau1933,Pekar1,Pekar2}. An electron moving in a solid polarizes its surroundings. The sufficiently strong carrier-boson coupling forms a bound state of the carrier and the cloud of bosonic excitations, lowering the energy of the ground state in comparison with the non-interacting system. The extended polaron concept deals with different kind of bosons including spin, charge, orbital or phonon fluctuations. However, initial formulation of the problem draws particular attention, since lattice vibrations are known or believed to play a significant role for many materials with extremely interesting properties. These include but not limited to high-temperature superconductors, ferromagnetic oxides, colossal magnetoresistive materials, conducting polymers, organic semiconductors, and molecular nanowires.

The central issue in the theory of polaron-phonon and exciton-phonon systems is a problem of transitions controlled by the parameters of the electron-phonon interaction (EPI). The history starts from the question of the phonon-induced self-trapping, which, as expected, should manifest itself as a point of non-analyticity in the energy of the ground state depending on coupling parameter. To continue, it is necessary to clarify that in general there are two types of the carrier-boson couplings. The first one modifies the on-site potential of the particle and depends only on boson momentum {\bf {q}}. The second one modulates intersite parameters and depends also on quasiparticle momentum {\bf {k}}. In a real space formulation these mechanisms correspond to diagonal and off-diagonal interactions, respectively. It has been shown previously that only smooth crossovers are allowed in the model with diagonal coupling \cite{RevModPhys.63.63}. Sharp polaron transitions have been found in the models where the off-diagonal interaction or the sum of diagonal and off-diagonal ones are taken into account \cite{PhysRevB.78.214301,PhysRevLett.105.266605,Sboychakov_2010,PhysRevLett.110.223002,PhysRevB.89.144508,PhysRevB.93.035130,PhysRevB.95.035117,NatComm8.2267,SciRep7.1169.2017,PhysRevB.101.134301}. All transitions found are controlled by the parameters of the electron-phonon interaction. However their origin is not necessarily the result of the transition between the large and small polaron states. This may occur, for example, due to the multi-band contributions \cite{PhysRevB.93.035130} or even due to the level crossing in polaron spectra \cite{NatComm8.2267}.

The crucial importance of the off-diagonal electron-phonon interaction in models with short-range electron-phonon coupling~\cite{PhysRevLett.117.206404,PhysRevResearch.2.023013} and even its physical consequences in the properties of high-temperature superconducting systems~\cite{PhysRevLett.92.146403,PhysRevLett.93.117004,PhysRevB.69.144520,PhysRevB.81.155116,Eremin2014} have been indicated earlier by various researches. Sharp polaron transitions, non-trivial topological effects \cite{PhysRevB.93.035130,PhysRevB.100.075126}, competing order \cite{Sboychakov_2010,PhysRevB.81.155116} and unconventional criticality~\cite{PhysRevResearch.2.023013}, all these interesting phenomena revealed in different models with off-diagonal electron-phonon coupling, arise the question: what can we expect in a real system where lattice contributions and other relevant interactions compete with one another? Here we consider a realistic two-dimensional multiband pd-model of cuprates, the parent compounds of high-temperature superconductors. These materials, like other transition metal oxides, are characterized by strong electron correlations, that cause a Mott-Hubbard insulating ground state of the undoped systems. The strong electron-phonon interaction, natural for these ionic compounds, is also well documented experimentally \cite{doi:10.1002/pssb.200404951,doi.org/10.1038/nature04973}. Even more, the features of the ARPES and optical conductivity spectra, which have been explained using the Franck-Condon broadening concept, indicate the polaronic behavior of the correlated carriers at low doping \cite{PhysRevLett.93.267002,RevModPhys.77.721,Mishchenko:2009}.

To capture the problem of the crossover from a rather mobile large polaron to a quasi-immobile small polaron we extend the pd-model with electron-phonon contributions coming from the modulation of both the on-site energy of electrons and their hopping parameter. Then to provide a proper description of this many-body task in the limit of strong electron correlations we apply the polaronic version of the generalized tight-binding (pGTB) method \cite{Ovchinnikov2012,PhysRevB.92.155143}. This approach is a kind of cluster perturbation theory; it combines the exact diagonalization of the unit cell cluster Hamiltonian and perturbation treatment of the interclaster contributions. The formulation of the pGTB method allows one to take into account such essential features of the material as complex unit cell, doping level, band structure and its realistic parameters obtained from \textit{ab initio} calculation. 

Progress in the polaron theory has recently been marked by new computational techniques and new theoretical approaches. These are numerical many-body methods, like the exact diagonalization procedure \cite{FehskeTrugman}, the continuous-time path-integral quantum Monte Carlo \cite{PhysRevB.72.035122,Kornilovitch} and diagrammatic Monte Carlo \cite{Mishchenko:2005,PhysRevLett.113.166402} simulations or the adaptive time-dependent density-matrix renormalization group \cite{PhysRevLett.93.076401,PhysRevB.78.035209}. These approaches do not involve any approximation, but they are limited by computational resources and have been mainly applied to idealized models. The atomistic calculations of polarons are also developing very extensively, moving towards fully \textit{ab-initio} consideration of the problem \cite{Verdi2017,PhysRevLett.122.246403}. However, for strongly correlated systems the calculations from ``first principles'' still face known difficulties related to the problem of approximation of the exchange-correlation potential.

If the Coulomb energy $U$ is much higher than the kinetic one $t$, the formation of high and low Hubbard bands and the spectral weight redistribution between them impact the properties of the bound electron-phonon states, and these effects should be carefully taken into account. To that end, the reported study of the polaron charge carries in the realistic model with strong electron-electron and electron-pnonon interactions employs the natural in the atomic limit $U \gg t$ approaches that  exactly treats all local contributions and properly account for the Hubbard character of quasiparticles in correlated system. With the pGTB method we consider the specific features of the polaron band structure and the phonon spectral function of the polaron in the limit of strong electron correlations and discuss their origin.

The paper is organized as follows. In Section~\ref{sec_formalism} we give the main ideas of the pGTB method and discuss the quasiparticle description of the bound electrons and phonons in strongly correlated system. In Section~\ref{sec_model} we present the model under study. Then theoretical details of the approach are provided. In Section~\ref{sec_Transformation} the Hamiltonian is transformed by introducing orthogonal Wannier functions centered on the unit cell site. In Section~\ref{sec_GF} the presentation of the Hubbard operators is applied. Equations for the Green's function of electron and phonon excitations in the system with strong electron-electron and electron-phonon interactions are obtained in Section~\ref{sec_DEq}. The computational details are given in Section~\ref{sec_Computer}. The evolution of the polaronic band structure is considered as a function of diagonal and off-diagonal electron-phonon coupling parameters in Section~\ref{sec_BandStrc}. The phonon spectral function of polaron is analyzed in Section~\ref{sec_PhSF}. In Sections~\ref{sec_Conclusion} discussion and conclusions are presented.

\section{Quasiparticles in the atomic limit \label{sec_formalism}}
The basic idea of pGTB approach is to divide the infinite lattice into identical unit cell. The corresponding new form of the Hamiltonian allows one to find exact Green's functions of a cluster and then to consider intercluster hoppings and interactions using a perturbation theory. The total Hamiltonian in the most general formulation reads
\begin{eqnarray}
 \label{H_tot_start}
H = {H_{el}} + {H_{ph}} + {H_{epi}},
\end{eqnarray}
where $ H_{el}$ describes the electronic structure of correlated system, $ H_{ph}$ is the free phonons term, and $ H_{epi}$ is the contribution of the electron-phonon interaction. For now, we skip the solution to concomitant problem of non-orthogonality of molecular orbitals of adjacent cells, which is outlined in the Section~\ref{sec_Transformation}, and rewrite the total Hamiltonian as a sum of $H_c$ intracell and  $H_{cc}$ intercell contributions:
\begin{eqnarray}
 \label{HcHcc}
H = {H_c} + {H_{cc}},{\quad}{H_c} = \sum\limits_{\bf{g}} {{H_{\bf{g}}}} ,{\quad}{H_{cc}} = \sum\limits_{ {{\bf{g}}\ne{\bf{g}}'} } {{H_{{\bf{gg}}'}}}.
\end{eqnarray}
The exact diagonalization of $H_{\bf{g}}$ gives us the complete set of eigenstates $\left| {p} \right\rangle $ with their energies $E_{p}$. Here, state index $p$ includes information about charge carries per unit cell $n$ and all other quantum numbers like spin, orbital moment, etc. The procedure is performed for each subspace $n$ of the Hilbert space, corresponding to relevant configurations of the separate ion in the crystal field. For example, for cuprate oxides these are mostly $ {d^9} {p^6}$ and ${d^{10}} {p^5} $ with 1 hole per unit cell, $n_h=1$, a hole vacuum ${d^{10}} {p^6}$ with $n_h=0$, and two-hole configurations ${d^9} {p^5}$, ${d^8} {p^6}$, and ${d^{10}} {p^4}$ with $n_h=2$.

If free phonons and electron-phonon interaction are absent in the total Hamiltonian, then eigenstates $\left| {p} \right\rangle$ are multielectron. Strong electron-lattice coupling leads to the multiplication of each electronic state $\left| {p} \right\rangle$ into a set $\left| {{p}_{i}} \right\rangle $ of polaronic states. 
These multielectron and multiphonon states can be written as
\begin{eqnarray}
 \label{pstates}
\left| {p_i} \right\rangle  = \sum\limits_\varsigma  {\sum\limits_{\nu  = 0}^{{N_{\max }}} {c_{i\nu }^\varsigma \left| \varsigma  \right\rangle \left| \nu  \right\rangle } }.
\end{eqnarray}
Here $\left| \varsigma  \right\rangle $ denotes a basic wave function of the multielectron state,   $\left| \nu  \right\rangle $ is a multiphonon state with number of phonons $n_{ph} = \nu$, this state results from the $\nu $ action of a phonon creation operator on the vacuum state of a harmonic oscillator, $N_{max}$ is a number of phonons involved in the formation of the polaronic state at a given strength of EPI, and $c_{i\nu }^\varsigma $ are coefficients that determine the structure of the polaronic state. 

Thus, there are many possible states for each subspace of a Hilbert space. Let us consider the state with the lowest energy in the sector with $n_h=1$, which is occupied at zero temperature and zero doping. Excitations from this ground state term to any others in the same subspace are accompanied by changes in the multiphonon structure of the polarons. However, the number of holes in the initial and final states is preserved. So, these are local excitonic Bose quasiparticles, the excitation energy of which is determined by the energy difference of the participating states. Interactions between different unit cells lead to a momentum dependence of the quasiparticle excitation energies and broaden out them into the bands forming the spectrum of phonons bounded with correlated electrons. Similarly, one can consider the Fermi type quasiparticle excitations bands, which correspond to the processes of creation or annihilation of an electron surrounded by a phonon cloud. The superposition of these Fermi or Bose type Hubbard quasiparticle excitations with their own spectral weights determines, respectively, the electron or phonon spectral properties of polaron in strongly correlated system with strong electron-phonon interaction.

The intercluster contributions in the present paper describe, besides mean-field terms, the interaction of charge carriers with spin fluctuations and inelastic scattering of quasiparticles by lattice vibrations. We apply here the so-called mode-coupling approximation for the lowest order self-energy which is equivalent to the non-crossing approximation in the diagram technique \cite{Tserkovnikov,Plakida2012}. Based on the exact intracell solutions that provide an adequate description of quasiparticles in the atomic limit, we self-consistently compute the system of Dyson equations for the electron and phonon Green's functions together with equations for the chemical potential. The approach takes into account the underlying features of the band structure formation in correlated system, such as, spectral weight redistribution between Hubbard quasiparticles subbands, and can be treated as the first reasonable approximation to the problem.

It should be added that intensity of the quasiparticle excitations is determined by the overlapping of matrix elements of the participating states and their occupation numbers. Thus, spectral weight redistribution between quasiparticle bands depends on doping, temperature, and parameters of electron-phonon coupling and Coloumb interactions, as well as on the momentum due to non-local contributions of the interactions taken into account. Such mechanism of mutual influence of electron-electron and electron-phonon interactions is characteristic for correlated compounds. Below we consider how this interplay affects the observable renormalizations of the electron band structure and phonon spectral functions.

\section{The model \label{sec_model}}
We define the total Hamiltonian (Eq.~\ref{H_tot_start}) on a two-dimensional (2D) square lattice with an electron contribution $H_{el}$  given by a realistic multiband p-d model. In a hole representation usually used for cuprates, it reads
\begin{eqnarray}
 \label{H_el}
{H_{el}} = \sum\limits_{{\bf{g}},l,\sigma } {\left[ { \sum\limits_{\alpha }{\left( {\varepsilon _\alpha }n_{{{\bf{g}}_\alpha },\sigma }^\alpha  + \frac{1}{2}{U_\alpha}n_{{{\bf{g}}_\alpha },\sigma }^\alpha n_{{{\bf{g}}_\alpha },\bar \sigma }^\alpha \right)} + } \right.} \nonumber \\
 + \sum\limits_{{\bf{g}}',l' } {{P_{pp}}{t_{pp}}\left( {p_{{{\bf{g}}_l},\sigma }^\dag {p_{{{\bf{g}}'_{l'}},\sigma }} + {\rm{H}.c.}} \right)}  + \nonumber \\
\left. { + {P_{pd}}{t_{pd}}\left( {d_{{\bf{g}},\sigma }^\dag {p_{{{\bf{g}}_l},\sigma }} + {\rm{H.c.}}} \right) + \sum\limits_{\sigma '} {{V_{pd}}n_{{{\bf{g}}_l},\sigma }^pn_{{\bf{g}},\sigma '}^d} } \right].
\end{eqnarray}
Here operators $d_{\bf{g},\sigma}^\dag $ and $p_{{\bf{g}}_l,\sigma}^\dag $ create a hole with spin $\sigma$ on copper and oxygen plane atomic orbitals at positions indicated by vectors ${\bf{g}}$ or ${\bf{g}}_{l}$, respectively, index $l$ enumerates oxygen atoms in the unit cell at site $\bf{g}$, $ \bar \sigma  =  - \sigma $, $n_{{{\bf{g}}_\alpha },\sigma }^\alpha$ is the hole number operators and ${\bf{g}}_{\alpha} = \bf {g}$ for $\alpha= d$ and $ {\bf{g}}_{\alpha} = {\bf{g}}_{l} $ for $\alpha = p$, $ \varepsilon_{\alpha} = \left( \varepsilon_ {\alpha, 0} - \mu \right)$ is a value of hole energy level with respect to the chemical potential $\mu$, $t_{pp}$ and $t_ {pd}$ are the hopping parameters. The phase factors $P_{pp}$ and $P_{pd}$ are equal to $1$ or $-1$, depending on whether orbitals with real wave functions have the same or opposite sign in the overlap region.

The free phonon Hamiltonian $H_{ph}$ from Eq.~\ref{H_tot_start} describes harmonic optical vibrations
\begin{eqnarray}
 \label{H_ph}
H_{ph}= \sum\limits_{\bf{q}} {\hbar {\omega _0}\left( {f_{\bf{q}}^\dag {f_{\bf{q}}} + {\frac{1}{2}}} \right)} = \sum\limits_{\bf{g}} {\hbar {\omega _0}\left( {f_{\bf{g}}^\dag {f_{\bf{g}}} + {\frac{1}{2}}} \right)}.
\end{eqnarray}
Here $f_{\bf{q}}^\dag $ is an operator of creation of phonon with momentum $\bf{q}$ and $f_{\bf{g}}^\dag $ is its Fourier transform that corresponds to the superposition of oxygen displacements in the unit cell centered at site $\bf{g}$. The contribution of much heavier copper atoms are omitted. To simplify the problem we consider~\cite{pGTB2018} only phonons of one planar oxygen bond-stretching breathing mode with energy $\omega_0$. We believe that in the absence of the electron-phonon interaction this mode is dispersionless. Due to the interaction it acquires the dispersion (Fig.~\ref{fig:f7}), which corresponds to the experimental one \cite{doi:10.1002/pssb.200404951,PhysRevB.60.R15039}.

The interaction of electrons to this phonon mode modulates the copper on-site energy and the copper-oxygen hopping parameter and is characterized by strong electron-phonon coupling strength in underdoped materials \cite{doi:10.1002/pssb.200404951,PhysRevB.60.R15039,PhysRevLett.82.628}. The linear in atomic displacements part is defined as
\begin{eqnarray}
 \label{H_epi}
{H_{epi}} &=& \sum\limits_{{\bf{g}},\sigma } {{M_d}\left( {f_{\bf{g}}^\dag  + {f_{\bf{g}}}} \right)d_{{\bf{g}},\sigma }^\dag {d_{{\bf{g}},\sigma }}}  + \nonumber \\
 &+& \sum\limits_{{\bf{g}},l,\sigma } {{M_{pd}}{P_{pd}}\left( {f_{\bf{g}}^\dag  + {f_{\bf{g}}}} \right)\left( {d_{{\bf{g}},\sigma }^\dag {p_{{\bf{g}}_{l},\sigma }} + {\rm{H.c.}}} \right)},
\end{eqnarray}
where $M_d$ and $M_{pd}$ are diagonal and off-diagonal values of electron-phonon interaction, respectively.

We use the set of parameters which corresponds to the single-layer copper oxide system La$_2$CuO$_4$: ${\varepsilon _{d,0}} = 0$,  $\varepsilon _{p,0} = 1.5$,  ${t_{pp}} = 0.86$, $t_{pd} = 1.36$, ${U_d} = 9$,  ${U_p} = 4$, and ${V_{pd}} = 1.5$, (all in eV). The values of the hoppings and single-electron-energies have been calculated earlier within the \textit{ab initio} local density approximation using Wannier function projection technique\cite{PhysRevB.72.165104}. The values of the Coulomb parameters are taken from the fitting to experimental ARPES data \cite{PhysRevLett.74.964}. The bare phonon frequency $ \omega_0$  is chosen equal to $90 $~meV. The electron-phonon coupling strength is characterized by dimensionless parameters: ${\lambda _{d\left( {pd} \right)}} = {{M_{d\left( {pd} \right)}^2} \mathord{\left/  {\vphantom {{M_{d\left( {pd} \right)}^2} {W\hbar {\omega _0}}}} \right. \kern-\nulldelimiterspace} {W\hbar {\omega _0}}}$, where $W$ is usually assumed to be the width of the free electron band. We calculate the value of $W$ for the system without electron-phonon interaction, but emphasize that in our problem it corresponds to the width of the occupied valence band of correlated electrons. Therefore, for the above set of parameters, we get $W = 2.15$~eV (Figs.~\ref{fig:f2}~(a)).

\section{The transformation of the Hamiltonian \label{sec_Transformation}}
To divide the Hamiltonian from Eq.~\ref{H_tot_start} on intracell and intercell contributions we transform all operators relating to oxygen sites to the basis centered on copper and orthogonal in different cells. The corresponding canonical transformation for oxygen orbitals can be found by inspection of Eq.~\ref{H_el} presented in the reciprocal ${\bf{q}}$~space \cite{PhysRevLett.63.1288}. Note that for operators not centered on the lattice sites, the Fourier transform obeys the following rule:
\begin{eqnarray}
 \label{FourierTr} 
{p_{{\bf{g}} \pm {{\bf{r}}_l},\sigma }} = \frac{1}{{\sqrt N }}\sum\limits_{\bf{q}} {{p_{l,{\bf{q}},\sigma }}\exp \left[ { - i{\bf{q}}\left( {{\bf{g}} \pm {{\bf{r}}_l}} \right)} \right]},
\end{eqnarray}
where ${{\bf{r}}_l} = {{{{\bf{a}}_l}} \mathord{\left/ {\vphantom {{{{\bf{a}}_l}} 2}} \right. \kern-\nulldelimiterspace} 2}$ and ${{{\bf{a}}_l}}$ is primitive vector of tetragonal lattice with $l=x,y$. The transformation of the fermionic operators ${p_{x,{\bf{q}},\sigma}}$ and $p_{y,{\bf{q}},\sigma}$ to the Wannier functions $b_{{\bf{q}},\sigma }$ and $a_{{\bf{q}},\sigma }$ is described for the pd-model in details elsewhere \cite{PhysRevB.53.8774, Gav2000}. The same procedure has been proposed for the bosonic operators related to oxygen sites \cite{PhysRevB.59.14697}. We have developed the idea \cite{PhysRevB.92.155143} and modified its formulation \cite{pGTB2018}. 

The relation of the phonon field operators to the phonon displacement operator for one dispersionless phonon mode reads:
\begin{eqnarray}
 \label{QPhField} 
{Q_{\bf{g}}} = \sqrt {\frac{\hbar }{{2m{\omega _0}}}} \left( {f_{\bf{g}}^\dag  + {f_{\bf{g}}}} \right).
\end{eqnarray}
At the same time, the normal coordinate of the planar oxygen breathing vibrations corresponds to the following superposition of the displacements of oxygen atoms:
\begin{eqnarray}
 \label{QUF} 
{Q_{\bf{g}}} = \frac{1}{2}\sum\limits_l {\left( {{U_{{\bf{g}} + {{{{\bf{a}}_l}} \mathord{\left/
 {\vphantom {{{{\bf{a}}_l}} 2}} \right.
 \kern-\nulldelimiterspace} 2}}} - {U_{{\bf{g}} - {{{{\bf{a}}_l}} \mathord{\left/
 {\vphantom {{{{\bf{a}}_l}} 2}} \right.
 \kern-\nulldelimiterspace} 2}}}} \right)}  = \nonumber \\
 = \frac{1}{{\sqrt N }}\sum\limits_{\bf{q}} {\left( { - i} \right){{\mathop{\rm e}\nolimits} ^{ - i{\bf{qg}}}}\left( {{s_{x,{\bf{q}}}}{U_{x,{\bf{q}}}} + {s_{y,{\bf{q}}}}{U_{y,{\bf{q}}}}} \right)},
\end{eqnarray}
where ${s_{x\left( y \right),{\bf{q}}}} = \sin \left( {{{{q_{x\left( y \right)}}{a_{x\left( y \right)}}} \mathord{\left/
 {\vphantom {{{q_{x\left( y \right)}}{a_{x\left( y \right)}}} 2}} \right.
 \kern-\nulldelimiterspace} 2}} \right)$.
It is obviously now that we can introduce ``canonical'' bosons:
\begin{eqnarray}
 \label{CanonicalTr}
\left( {\begin{array}{*{20}{c}}
{{A_{\bf{q}}}}\\
{{B_{\bf{q}}}}
\end{array}} \right) = \left( {\begin{array}{*{20}{c}}
{{{ - i{s_{x,{\bf{q}}}}} \mathord{\left/
 {\vphantom {{ - i{s_{x,{\bf{q}}}}} {{\mu _{\bf{q}}}}}} \right.
 \kern-\nulldelimiterspace} {{\mu _{\bf{q}}}}}}&{{{ - i{s_{y,{\bf{q}}}}} \mathord{\left/
 {\vphantom {{ - i{s_{y,{\bf{q}}}}} {{\mu _{\bf{q}}}}}} \right.
 \kern-\nulldelimiterspace} {{\mu _{\bf{q}}}}}}\\
{{{ - i{s_{y,{\bf{q}}}}} \mathord{\left/
 {\vphantom {{ - i{s_{y,{\bf{q}}}}} {{\mu _{\bf{q}}}}}} \right.
 \kern-\nulldelimiterspace} {{\mu _{\bf{q}}}}}}&{{{ + i{s_{x,{\bf{q}}}}} \mathord{\left/
 {\vphantom {{ + i{s_{x,{\bf{q}}}}} {{\mu _{\bf{q}}}}}} \right.
 \kern-\nulldelimiterspace} {{\mu _{\bf{q}}}}}}
\end{array}} \right)\left( {\begin{array}{*{20}{c}}
{{{\xi ^{ - 1}}U_{x,{\bf{q}}}}}\\
{{{\xi ^{ - 1}}U_{y,{\bf{q}}}}}
\end{array}} \right)
\end{eqnarray}
Here $\xi  = \sqrt {{\hbar  \mathord{\left/  {\vphantom {\hbar  {2m{\omega _0}}}} \right. \kern-\nulldelimiterspace} {2m{\omega _0}}}} $ and ${\mu _{\bf{q}}} = \sqrt {s_{x,{\bf{q}}}^2 + s_{y,{\bf{q}}}^2} $ is the coefficient of the transformation. This value and other coefficients resulting from the canonical transformations cause the nonlocal nature of the parameters in the Wannier representation. 
With transformed fermionic and bosonic operators the total Hamiltonian can be written in the form of Eqs.~\ref{HcHcc}: 
\begin{align}
 \label{cTransfH}
   {{H}_{c}}&=\sum\limits_{\mathbf{g},\sigma }{\left[ \sum\limits_{\beta }{\left( {{\varepsilon }_{\beta }}n_{\mathbf{g},\sigma }^{\beta }+\frac{1}{2}{{U}_{\beta }}n_{\mathbf{g},\sigma }^{\beta }n_{\mathbf{g},\bar{\sigma }}^{\beta } \right)} \right.+} \\  \nonumber
 & +t_{00}^{bd}\left( d_{\mathbf{g},\sigma }^{\dagger }{{b}_{\mathbf{g},\sigma }}+\operatorname{Hc} \right)+\sum\limits_{\sigma '}{{{V}_{bd}}n_{\mathbf{g},\sigma }^{d}n_{\mathbf{g},\sigma '}^{b}}+ \\  \nonumber
 & \left. +M_{00}^{d}\varphi _{\mathbf{g}}^{A}n_{\mathbf{g},\sigma }^{d}+M_{000}^{bd}\varphi _{\mathbf{g}}^{A}\left( d_{\mathbf{g},\sigma }^{\dagger }{{b}_{\mathbf{g},\sigma }}+\operatorname{Hc} \right) \right] \\ \nonumber
 & +\sum\limits_{\mathbf{g}}{h{{\omega }_{0}}\left( A_{\mathbf{g}}^{\dagger }{{A}_{\mathbf{g}}}+\frac{1}{2} \right)},\\
\label{ccTransfH}
{H_{cc}} &= \sum\limits_{ {{\bf{g}} \ne {\bf{g}}'} ,\sigma } {\left[ {t_{{\bf{gg}}'}^{bb}b_{{\bf{g}},\sigma }^\dag {b_{{\bf{g}}',\sigma }} + t_{{\bf{gg}}'}^{bd}\left( {d_{{\bf{g}},\sigma }^\dag {b_{{\bf{g}}',\sigma }} + {\rm{H}}{\rm{.c}}{\rm{.}}} \right) + } \right.} \nonumber \\
& + M_{{\bf{gg}}'}^d\varphi _{\bf{g}}^A n_{{\bf{g}}',\sigma }^d + \sum\limits_{\bf{h}} {M_{{\bf{g}\bf{g}'\bf{h}}}^{bd}\varphi _{\bf{g}}^A} \left. {\left( {d_{{\bf{g}}',\sigma }^\dag {b_{{\bf{h}},\sigma }} + {\mathop{\rm H.c.}\nolimits}} \right)} \right]
\end{align}
where $\beta=b,d$ is the orbital index, ${\varepsilon _b} = {\varepsilon _p} - 2{t_{pp}}{\nu _{00}}$, ${U_b} = {U_p}{\Psi _{0000}}$, and ${V_{bd}} = {V_{pd}}{\Phi _{000}}$, $t_{{\bf{gg}}'}^{bd} =  - 2{t_{pd}}{\mu _{{\bf{gg}}'}}$, $t_{{\bf{gg}}'}^{bb} =  - 2{t_{pp}}{\nu _{{\bf{gg}}'}},M_{{\bf{gg}}'}^d = {M_d}{\mu _{{\bf{gg}}'}}$, $M_{{\bf{g}\bf{g}'\bf{h}}}^{bd} =  - 2{M_{pd}}{\mu _{{\bf{gg}}'}}{\mu _{{\bf{hg}}'}}$ are renormalized parameters of the Hamiltonian given by Eqs.~(\ref{H_el},\ref{H_ph},\ref{H_epi}). The intercell contributions from $U_p$ and $V_{pd}$ interactions are omitted here due to their smallness. The coefficients ${\mu _{{\bf{gg}}'}}$ and ${\nu _{{\bf{gg}}'}}$ are Fourier transforms of ${\mu _{\bf{q}}}$ and  ${\nu _{\bf{q}}} = {{4s_{x,{\bf{q}}}^2s_{y,{\bf{q}}}^2} \mathord{\left/ {\vphantom {{4s_{x,{\bf{q}}}^2s_{y,{\bf{q}}}^2} {\mu _{\bf{q}}^2}}} \right. \kern-\nulldelimiterspace} {\mu _{\bf{q}}^2}}$, respectively. The involved values of transformation coefficients at the matching sites are given by the set: $\Psi _{0000} \approx 0.21$, $\Phi _{000} \approx 0.92$, $\mu _{00} \approx 0.96$, and $\nu _{00} \approx 0.73$ \cite{PhysRevB.53.8774,Gav2000}. The phonon contribution is described by the operator ${\varphi _{\bf{g}}^A} = \left( {A_{\bf{g}}^\dag  + {A_{\bf{g}}}} \right)$. The operator $B_{\bf{q}}$ from Eq.~\ref{CanonicalTr}  is completely decouples from any other degrees of freedom and hence eliminated from the problem \cite{pGTB2018}. We also reduced the three-band p-d model to a two-band one, neglecting the symmetric orbital $a_{{\bf{q}},\sigma }$. The simulations show that it has an insignificantly small influence on the low-lying eigenstates of the system when electrons interact with longitudinal oxygen vibrations in the CuO-plane.

\section{The Hamiltonian and Green's Functions in X-Operator Representation \label{sec_GF}}
To take into account the spectral weight redistribution between quasiparticles in correlated system, we use below the representation of the Hubbard operators. Having the complete set of the multielectron and multiphonon eigenstates $\left| p \right\rangle$, defined by exact diagonalization of the Hamiltonian $H_c$ (Eq.~\ref{cTransfH}), any operator $O_{\bf{g}}$ can be presented as a linear combination of the Hubbard X operators: 
\begin{eqnarray}
 \label{O_inX}
O_{\bf{g}} = \sum\limits_{p,p'} {\left\langle p \right|O_{\bf{g}}\left| {p'} \right\rangle {X_{\bf{g}}^{p,p'}}}  = \sum\limits_{P} {{\gamma}_{O,P} {X_{\bf{g}}^{P}}},
\end{eqnarray}
where ${\gamma _{O,P}} = \left\langle {p\left| {{O_{\bf{g}}}} \right|p'} \right\rangle $ is a matrix element of transition from initial $\left| p' \right\rangle $ to final $\left| p \right\rangle $ states, i.e. for a pair of states $P = \left( {p,p'} \right)$, under the action of the operator $O$.
Then total Hamiltonian takes the simple form:
\begin{align}
\label{X_Hc}
&H_c = \sum\limits_{{\bf{g}},p} {{E_{p}}{X_{\bf{g}}^{p,p}}}, \\
&{H_{cc}} = \sum\limits_{ {{\bf{g}}\ne{\bf{g}}'\ne{\bf{h}}} } {\sum\limits_{P,Q} {\left( {T_{{\bf{gg}}'}^{PQ}{\delta _{{\bf{hg}}'}}\overset{\dagger }{\mathop{X_{\mathbf{g}}^{P}}}X_{{\bf{g}}'}^{Q} + } \right.} } \nonumber \\
\label{X_Hcc}
& + \sum\limits_N {\left( {M_{{\bf{gg}}'}^{NP}{\delta _{{\bf{hg}}'}}{\delta _{PQ}} + M_{{\bf{gg}}'{\bf{h}}}^{NPQ}} \right)\left. {X_{\bf{g}}^N\overset{\dagger }{\mathop{X_{\mathbf{g}'}^{P}}}X_{\bf{h}}^{Q}} \right)}, 
\end{align}
with coefficients as 
\begin{align*}
&T_{{\bf{gg}}'}^{PQ} = t_{{\bf{gg}}'}^{bb}\gamma _{b_\sigma ,P}^\dag {\gamma _{{b_\sigma },Q}} + t_{{\bf{gg}}'}^{bd}\Gamma _{PQ}^{bd}, \\
&M_{{\bf{gg}}'}^{NP} = M_{{\bf{gg}}'}^d\Gamma _N^A\gamma _{{d_\sigma },P}^\dag {\gamma _{{d_\sigma },P}},\  
M_{{\bf{gg}}'{\bf{h}}}^{NPQ } = M_{{\bf{ghg}}'}^{bd}\Gamma _N^A\Gamma _{PQ}^{bd}, \\
&\Gamma _{PQ}^{bd} = \gamma _{d_\sigma ,P}^\dag {\gamma _{{b_\sigma },Q}} + \gamma _{b_\sigma ,P}^\dag{\gamma _{{d_\sigma },Q}}, \  \Gamma _N^A = \gamma _{{A},N}^\dag  + {\gamma _{A,N}}.
\end{align*}
The cost of a simple presentation of the Hamiltonian in the form of Eqs.~\ref{X_Hc} and ~\ref{X_Hcc} is a complicated commutation relations for the Hubbard operators
\begin{eqnarray}
\label{Commut}
{{\left\{ X_{\mathbf{g}}^{p,p'},X_{\mathbf{g}'}^{q,q'} \right\}}_{\pm }}={{\delta }_{\mathbf{g},\mathbf{g}'}}\left( {{\delta }_{p',q}}X_{\mathbf{g}}^{p,q'}\pm {{\delta }_{q',p}}X_{\mathbf{g}}^{q,p'} \right).
\end{eqnarray}
The commutator is taken with a minus sign if both or one of the operators $X_{\mathbf{g}}^{p,p'}$ or $X_{\mathbf{g}'}^{q,q'}$ are quasibosonic one, so the difference of holes in the initial and final states for the given operator is even. If operators $X_{\mathbf{g}}^{p,p'}$ and $X_{\mathbf{g}'}^{q,q'}$ are quasifermionic, that is $\left( {{n}_{p}}-{{n}_{p'}} \right)$ and $\left( {{n}_{q}}-{{n}_{q'}} \right)$ are odd, then commutator is taken with a plus sign. 

Now we define the Green's function describing phonon excitations of polarons. The description of boson excitations in the Hubbard representation was demonstrated previously for spin waves and phonons~\cite{Zaitsev75}. In general, the phonon Green's function for the dispersionless mode has the form
\begin{eqnarray}
 \label{PhGrTot}
{{D}_{\mathbf{gg}'}}\left( t,t' \right) &=&-i\left\langle T{{\varphi }_{\mathbf{g}}}\left( t \right){{\varphi }_{\mathbf{g}'}}\left( t' \right) \right\rangle \\ \nonumber
&=& \left\langle \left\langle  {{\varphi }_{\mathbf{g}}}\left( t \right) | {{\varphi }_{\mathbf{g}'}}\left( t' \right) \right\rangle  \right\rangle,
\end{eqnarray}
where  $T$ is the time-ordering operator and the phonon field operator reads ${\varphi _{\bf{g}}} = \sqrt{\frac{\omega_0 }{2}} \left( {{A_{\bf{g}}} + A_{\bf{g}}^\dag } \right)$. Here and below the notations of Zubarev~\cite{Zubarev:1960} for the Green's function are used. According to Eq.~\ref{O_inX},
\begin{eqnarray}
\label{A_phononX}
{A_{\bf{g}}} = \sum\limits_{n,i,i'} {\left\langle {ni\left| {{A_{\bf{g}}}} \right|ni'} \right\rangle X_{\bf{g}}^{ni,ni'}}  = \sum\limits_{N_v} {{\gamma _{A,N_v}}X_{\bf{g}}^{N_v}}.
\end{eqnarray}
The indexes $N_v$ indicate all allowed transitions between pair of the eigenstates with the same number of holes, $N_v=\left( ni,ni' \right)$.
The function ${D_{{\bf{gg}}'}}\left( {t,t'} \right)$ can be presented now as a linear combination of the Green's functions of the Bose Hubbard quasiparticles
\begin{equation}
 \begin{aligned}
\label{PhGrX}
 {{D}_{\mathbf{gg}'}}\left( t,t' \right)={\frac{{{\omega }_{0}}}{2}}\\ 
  \sum\limits_{N_v,N_{v'}}{\left\{ {{\gamma }_{A,N_v}}{{\gamma }_{A,N_{v'}}} \left\langle  \left\langle \overset{{}}{\mathop{X_{\mathbf{g}}^{N_{v}}}}\,\left( t \right) \right. | \left. \overset{{} }{X}{}_{\mathbf{g}'}^{N_v'}\,\left( t' \right) \right\rangle  \right\rangle \right.+} \\ 
+{{\gamma }_{A,N_v}}\gamma _{{A},N_{v'}}^{\dagger } \left\langle \left\langle  X_{\mathbf{g}}^{{{N}_{v}}} \,\left( t \right) | \overset{\dagger }{X}{}_{\mathbf{g}'}^{{{N}_{v'}}} \,\left( t' \right) \right\rangle  \right\rangle + \\
 +\gamma _{{A},N_v}^{\dagger }{{\gamma }_{A,N_{v'}}}\left\langle  \left\langle \overset{\dagger }{X}{}_{\mathbf{g}}^{N_v}\,\left( t \right) \right. | \left. \overset{{}}{\mathop{X_{\mathbf{g}'}^{N_{v'}}}}\,\left( t' \right) \right\rangle  \right\rangle + \\ 
 \left. +\gamma _{{A},N_v}^{\dagger }\gamma _{{A},N_{v'}}^{\dagger }\left\langle  \left\langle \overset{\dagger }{X}{}_{\mathbf{g}}^{N_v}\,\left( t \right) \right. | \left. \overset{\dagger }{X}{}_{\mathbf{g}'}^{N_{v'}}\,\left( t' \right) \right\rangle  \right\rangle  \right\}. 
\end{aligned}
\end{equation}
To order the set of operators $\left\{ X_{\mathbf{g}}^{N_v} \right\}$ we introduce the multicomponent operator $\hat{X}_{\mathbf{g}}^{N_V}\equiv \hat{X}_{\mathbf{g}}^{ni,ni'}={{\left( X_{\mathbf{g}}^{{{N}_{1}}},X_{\mathbf{g}}^{{{N}_{2}}},...X_{\mathbf{g}}^{{{N}_{v}}},...X_{\mathbf{g}}^{{{N}_{V}}} \right)}^{T}}$, where $V$ is the maximum number of transitions $N$ taken into account. Then we continue with the matrix quasiparticle Green's function
\begin{eqnarray}
\label{PhGrQP}
{{\hat{D}}_{\mathbf{gg}'}}\left( t,t' \right)=
\left\langle {\left\langle {{\hat X_{\bf{g}}^{{N_V}}\left( t \right)}}
 \mathrel{\left | {\vphantom {{\hat X_{\bf{g}}^{{N_V}}\left( t \right)} {{{\left( {\hat X_{{\bf{g}}'}^{{N_V}}\left( {t'} \right)} \right)}^T}}}}
 \right. \kern-\nulldelimiterspace}
 {{{{\left( {\hat X_{{\bf{g}}'}^{{N_V}}\left( {t'} \right)} \right)}^T}}} \right\rangle } \right\rangle ,
\end{eqnarray}
which consists of four blocks corresponding to individual phonon excitations such as $A_\mathbf{g}A_{\mathbf{g}'}$, ${{A}_{\mathbf{g}}}A_{{\mathbf{{g}'}}}^{\dagger }$, $A_{\mathbf{g}}^{\dagger }{{A}_{\mathbf{g}'}}$, and $A_{\mathbf{g}}^{\dagger }A_{\mathbf{g}'}^{\dagger }$. The components $D_{\mathbf{gg}'}^{{{N}_{v}}{{N}_{{v}'}}}\left( t,t' \right)$ of the matrix ${{\hat{D}}_{\mathbf{gg}'}}\left( t,t' \right)$ define the quasiparticle Green's functions of the phonon excitations of polaron.

Performing a similar procedure for the single-hole retarded Green's function we get
\begin{equation}
\begin{aligned}
\label{ElGrX}
 & G_{\mathbf{gg}',\sigma}^{\beta \beta '}\left( t,t' \right)=
\left\langle {\left\langle {{{a_{\beta ,{\bf{g}},\sigma }}\left( t \right)}}
 \mathrel{\left | {\vphantom {{{a_{\beta ,{\bf{g}},\sigma }}\left( t \right)} {a_{\beta ',{\bf{g}}',\sigma '}^\dag \left( {t'} \right)}}}
 \right. \kern-\nulldelimiterspace}
 {{a_{\beta ',{\bf{g}}',\sigma '}^\dag \left( {t'} \right)}} \right\rangle } \right\rangle = \\
 & =\sum\limits_{P_h,P_{h'}}{{{\gamma }_{{{a}_{\beta ,\sigma }},P_h}}\gamma _{{{a}_{\beta '\sigma }},P_{h'}}^{\dagger }
\left\langle {\left\langle {{X_{{\bf{g}},\sigma }^{{P_h}}\left( t \right)}}
 \mathrel{\left | {\vphantom {{\overset{\dagger } {X}{}_{{\bf{g}},\sigma }^{{P_h}}\left( t \right)} {X_{{\bf{g}}',\sigma }^{{P_{h'}}}\left( {t'} \right)}}}
 \right. \kern-\nulldelimiterspace}
 {\overset{\dagger}{X}{}_{{\bf{g}}',\sigma }^{{P_{h'}}}\left( {t'} \right)} \right\rangle } \right\rangle }, 
\end{aligned}
\end{equation}
here $a_{\beta ,\mathbf{g},\sigma }^{\dagger }$ is the operator of creation of a hole with spin $\sigma$ at a lattice site $\mathbf{g}$ on the orbital $\beta$.
The indexes $P_h$ indicate all allowed transitions between pair of eigenstates with decreasing of the hole number in the final state by one. If the set of operators $\left\{X_{\mathbf{g}}^{P_h} \right\}$ is ordered by vector ${{\left( \hat{X}_{\mathbf{g}}^{P_H} \right)}^{\dagger }}=\left( \overset{\dagger }{X}{}_{\mathbf{g}}^{{{P}_{1}}},\overset{\dagger }{X}{}_{\mathbf{g}}^{{{P}_{2}}},...\overset{\dagger }{X}{}_{\mathbf{g}}^{{{P}_{h}}},...\overset{\dagger }{X}{}_{\mathbf{g}}^{{{P}_{H}}} \right)$, with $H$ being the maximum number of transitions, then
\begin{eqnarray}
\label{FermiGrF}
\hat{G}_{\mathbf{gg}',\sigma}\left( t,t' \right)=
\left\langle {\left\langle {{\hat X_{{\bf{g}},\sigma }^{{P_H}}\left( t \right)}}
 \mathrel{\left | {\vphantom {{\hat X_{{\bf{g}},\sigma }^{{P_H}}\left( t \right)} {{{\left( {\hat X_{{\bf{g}}',\sigma }^{{P_{H'}}}\left( {t'} \right)} \right)}^\dag }}}}
 \right. \kern-\nulldelimiterspace}
 {{{{\left( {\hat X_{{\bf{g}}',\sigma }^{{P_{H'}}}\left( {t'} \right)} \right)}^\dag }}} \right\rangle } \right\rangle ,
\end{eqnarray}
and components $G_{\mathbf{gg}',\sigma}^{{{P}_{h}}{{P}_{h'}}}\left( t,t' \right)$ of the matrix $\hat{G}_{\mathbf{gg}',\sigma}\left( t,t' \right)$ define the quasiparticle Green's functions of the hole excitations of polaron.

\section{Dyson Equation \label{sec_DEq}}
We apply the projection operator method in the equation of motion for the thermodynamic two-time Green's function to research the Hamiltonian given by Eqs.~\ref{X_Hc} and~\ref{X_Hcc}. The method enables to derive a Dyson-type equation for an arbitrary Green's function with a self-energy similar to the memory function in the Mori projection technique \cite{10.1143/PTP.33.423}. It has been successfully used before by various authors to study systems where strong correlations are essential \cite{Kuzemsky,Balucani2003,Mancini2004}. The general formulation of the method can be found elsewhere \cite{Tserkovnikov,Plakida2012}, so we skip these details here.

By sequential differentiating each Green's function from Eqs.~\ref{PhGrQP} and \ref{FermiGrF} with respect to time $t$ and $t'$ and using the projection procedure, we obtain the Dyson equations
\begin{eqnarray}
\label{DysonEl}
  & \hat{G}_{\mathbf{k},\sigma }^{-1}\left( \omega  \right)={{\left( \hat{G}_{\mathbf{k},\sigma }^{0}\left( \omega  \right) \right)}^{-1}}-{{{\hat{\Sigma}}}_{\mathbf{k},\sigma }}\left( \omega  \right), \\
\label{DysonPh}
 & {\hat{D}_{\mathbf{q}}^{-1}}\left( \omega  \right)={{\left( {\hat{D}_{\mathbf{q}}^{0}}\left( \omega  \right) \right)}^{-1}}-\hat{\Pi }_{\mathbf{q}}\left( \omega  \right),
\end{eqnarray}
where $\hat{G}_{\mathbf{k},\sigma }\left( \omega  \right)$ and $\hat{D}_{\mathbf{q}}\left( \omega  \right)$ are the Fourier transforms of the matrix Green's functions $\hat{G}_{\mathbf{gg}',\sigma}\left( t,t' \right)$ and  $\hat{D}_{\mathbf{gg}'}\left( t,t' \right)$, respectively. It should be emphasized that the components of matrix ${\hat{D}_{\mathbf{q}}}\left( \omega  \right)$ describe the individual quasiparticle phonon excitations. The well-known form of the Dyson equation for the phonon Green's function from Eq.~\ref{PhGrTot} corresponds to a superposition of these excitations (Eq.~\ref{PhGrX}). The poles of the zero-order Green's functions $\hat{G}_{\mathbf{k}, \sigma }^{0}\left( \omega  \right)$ and ${\hat{D}_{\mathbf{q}}^{0}}\left( \omega  \right)$ define the excitation spectrum in the generalized mean-field approximation, for details see, for example~\cite{Plakida2012}. With the total Hamiltonian $H_c+H_{cc}$ the function $\hat{G}_{\mathbf{k}, \sigma }^{0}\left( \omega  \right)$ reads
\begin{eqnarray}
\label{zeroEl}
\hat{G}_{\mathbf{k},\sigma }^{0}\left( \omega  \right)={{\left( \omega {{{\hat{\tau }}}_{0,H}}-{{{\hat{\varepsilon }}}_{\mathbf{k},\sigma }} \right)}^{-1}}\hat{F}_H.
\end{eqnarray}
Here ${{{\hat{\tau }}}_{0,H}}$ is the $ H \times H $ unit matrix and $\hat{F}_H$ is the diagonal matrix of the occupation numbers of holes with elements defined as $F_{H}^{{{P}_{h,}}{{P}_{h'}}}=\left\langle {{\left\{ X_{\mathbf{g}}^{{{P}_{h}}},\overset{\dagger }{X}{}_{\mathbf{g}}^{{{P}_{h'}}}\, \right\}}_{+}} \right\rangle $, where the anomalous averages of the off-diagonal type are neglected. The quasiparticle spectrum is given by the matrix ${{{\hat{\varepsilon }}}_{\mathbf{k},\sigma }}$ with components $\varepsilon _{\mathbf{k},\sigma}^{{{P}_{h}}{{P}_{h'}}}$, which can be written as  
\begin{eqnarray}
\label{elspectr}
\varepsilon _{\mathbf{k},\sigma}^{{{P}_{h}}{{P}_{h'}}}=F_{H}^{{{P}_{h}}}F_{H}^{{{P}_{h'}}}T_{\mathbf{k}}^{{{P}_{h}}{{P}_{h'}}}\pm 
\sum\limits_{\mathbf{q}}{{{c}_{\mathbf{q}}}T_{\mathbf{k}-\mathbf{q}}^{{{P}_{h}}{{P}_{h'}}}}\\ \nonumber
+F_{H}^{{{P}_{h}}}{{E}_{{{P}_{h}}}}{{\delta }_{{{P}_{h}},{{P}_{h'}}}}.
\end{eqnarray}
Equation~\ref{elspectr} describes the quasiparticle bands of the Fermi-type excitations of correlated holes coupled with phonons. The sign (minus) plus is taken if the transitions ${P}_{h}$ and ${P}_{h'}$ correspond to the same (different) subspaces of the Hilbert space, ${\delta }_{{{P}_{h}},{{P}_{h'}}}$ is the Kronecker delta-function and it gives zero if ${h} \ne {h'}$.  We simplified the notation of the diagonal components of the matrix $\hat{F}_H$, so $F_{H}^{{{P}_{h}}{{P}_{h}}} = F_{H}^{{{P}_{h}}} = \left\langle { {X}_{\mathbf{g}}^{p,p} + {X}_{\mathbf{g}}^{p',p'} } \right\rangle$ for index ${P}_{h} = \left({p,p'}\right)$. The local quasiparticle excitation energy for the same index has the form ${E}_{{P}_{h}} = \left( {E_{p'} - E_{p}} \right)$. Fourier transform of the static spin correlation function is given by ${c_{\bf{q}}} = \sum\limits_{\left( {{\bf{f}} - {\bf{f}}'} \right)} {{c_{{\bf{ff}}'}}\exp \left[ { - {\bf{q}}\left( {{\bf{f}} - {\bf{f}}'} \right)} \right]} $ and ${c_{{\bf{ff}}'}} = 3\left\langle {S_{\bf{f}}^zS_{{\bf{f}}'}^z} \right\rangle$. This function has been calculated in Ref.~\cite{Korshunov2007} with the given above parameters of the LSCO system. It characterizes the spin liquid properties of the underdoped cuprates with short-range antiferromagnetic order and is concentration dependent.

The zero-order matrix function ${\hat{D}_{\mathbf{q}}^{0}}\left( \omega  \right)$ of the quasiparticle individual phonon excitations in the generalized mean-field approximation with the total Hamiltonian has the form
\begin{equation}
\label{phspectr}
{{{\hat{D}}}^{0}}\left( \mathbf{q},\omega  \right)={{\left( \omega {{{\hat{\tau }}}_{0,V}}-{{{\hat{\Omega }}}_{\mathbf{q} }} \right)}^{-1}}\hat{F}_{V}.
\end{equation}
Here, $\hat{F}_{V}$ is the matrix of the occupation numbers and its components are $F_{V}^{{{N}_{v,}}{{N}_{v'}}}=\left\langle {\left\{ X_{\mathbf{g}}^{{{N}_{v}}},\overset{\dagger }{X}{}_{\mathbf{g}}^{{{N}_{v'}}}, \right\}_{-}} \right\rangle $. The energy of the quasiparticle excitations is given by the matrix ${{{\hat{\Omega }}}_{\mathbf{q} }}$ with the components ${\Omega }_{\mathbf{q}}^{{N_v}{N_{v'}}} = E_{N_v} + \Delta E_{\mathbf{q}}^{{{N_v}{N_{v'}}}}$, where $E_{N_v} = E_{ni'} - E_{ni}$ and $\Delta E_{\mathbf{q}}^{{N_v}{N_{v'}}}$ is the contribution caused by non-local part of the Hamiltonian $H_{cc}$, for diagonal components ${\Omega }_{\mathbf{q}}^{{N_v}{N_{v}}}$ we use below the simple notation ${\Omega }_{\mathbf{q}}^{N_v}$. As an example, we give the expression for the phonon Green's function neglecting inelastic quasiparticle scattering, in accordance with Eq.~\ref{PhGrX} it takes the form
\begin{equation}
\label{Dzero}
D_{\mathbf{q}}^{0}\left( \omega  \right)=\sum\limits_{{{N}_{v}}}{{{\left| {{\gamma }_{A,{{N}_{v}}}} \right|}^{2}}F_{V}^{{{N}_{v}}}\frac{{{\omega }_{0}} {\Omega }_{\mathbf{q}}^{N_v} }{{{\omega }^{2}- {\left( {{{\Omega }_{\mathbf{q}}^{N_v} } }\right)}^2}}}.
\end{equation}
The low-energy spectrum of phonon excitations of polarons is equidistant if electron-phonon interaction is absent in the total Hamiltonian. Indeed, in this case, the energy difference between neighboring polaron terms exactly corresponds to the original frequency $\omega_0$ in each sectors of the Hilbert space, moreover the matrix elements are nonzero only for transitions with a change in the number of the initial and final polaron states by one.
As a result, at zero temperature and zero doping we get ${\Omega }_{\mathbf{q}}^{N_v} = \omega_0$ and Eq.~\ref{Dzero} describes the Green's function of free phonons for the dispersionless mode
\begin{equation}
\label{FreePhononGF}
 D^{0}\left( \omega  \right) =\frac{\omega _{0}^{2}}{{{\omega }^{2}}-\omega _{0}^{2}}.
\end{equation}
Electron-phonon interaction renormalizes the frequency $\omega_0$ of the phonons coupled with electrons. The strong electron correlations lead to the dependence of the renormalized  spectrum on doping and temperature via the occupation number in spectral weight of Green's function of quasiparticle excitations.
  
The self-energy operators ${{{\hat{\Sigma}}}_{\mathbf{k},\sigma }}\left( \omega  \right)$ and $\hat{\Pi }_{\mathbf{q}}\left( \omega  \right)$ for the given quasiparticle Green's functions have been obtained within the non-crossing approximation. We consider inelastic electron-phonon scattering assuming that the renormalization of the electron-phonon interaction vertex may be neglected, therefore two-particle correlation functions are decoupled in the following way
\begin{equation}
\begin{aligned}
\label{NCA}
\left \langle { \overset{\dagger }{X}{}_{\mathbf{q}'}^{{{N}_{v}}} \left( {t'}\right) \overset{\dagger } {X}{}_{\mathbf{k}-\mathbf{q}',\sigma}^{{{P}_{h}}} \left( {t'}\right) {X_{\mathbf{k}-\mathbf{q},\sigma}^{{{P}_{h'}}}} \left( {t}\right) {X_{\mathbf{q}}^{{{N}_{v'}}}} \left( {t}\right)} \right \rangle \simeq \\ \nonumber
\simeq {\left \langle {\overset{\dagger }{X}{}_{\mathbf{q}'}^{{{N}_{v}}} \left( {t'}\right) {X_{\mathbf{q}}^{{{N}_{v'}}}}\left( {t}\right)} \right \rangle \left \langle {\overset{\dagger }{X}{}_{\mathbf{k}-\mathbf{q}',\sigma}^{{{P}_{h}}}\left( {t'}\right) {X_{\mathbf{k}-\mathbf{q},\sigma}^{{{P}_{h'}}}} \left( {t}\right) } \right \rangle }.
\end{aligned}
\end{equation}
Expressing the single-particle correlators in terms of the corresponding Green's functions, we get equations for the components of matrices of the self-energy ${{{\hat{\Sigma}}}_{\mathbf{k},\sigma }}\left( \omega  \right)$ and polarization operator $\hat{{\Pi}_\mathbf{q} }\left( \omega  \right)$
\begin{widetext}
\begin{subequations} \label{MopNCA}
\begin{align}
{\Sigma}_{\mathbf{k},\sigma }^{{{P}_{h}}{{P}_{h'}}}\left( \omega  \right)=F_{H}^{{{P}_{h}}}F_{H}^{{{P}_{h'}}}\sum\limits_{\mathbf{q},{{N}_{v}}}{{{\left| M_{\mathbf{q},\mathbf{k}-\mathbf{q}}^{bd}  \Gamma _{{{N}_{v}}}^{A} \right|}^{2}}} \times  \\ \nonumber
 \int{\int{\frac{dzd\Omega }{{{\pi }^{2}}}f_H\left( \omega ,z,\Omega  \right)}} \left[ \sum\limits_{{{P}_{h''}},{{P}_{h'''}}}{\Gamma _{{{P}_{h}}{{P}_{h''}}}^{bd}}\overset{\dagger }{\Gamma}{} _{{{P}_{h'}}{{P}_{h'''}}}^{bd}\,\operatorname{Im}{{\left\langle \left\langle  X_{\mathbf{q}}^{{{N}_{v}}} | \overset{\dagger }{X}{}_{\mathbf{q}}^{{{N}_{v}}} \right\rangle  \right\rangle }_{\Omega }}\,\operatorname{Im}{{\left\langle \left\langle  X_{\mathbf{k}-\mathbf{q}}^{{{P}_{h''}}} | \overset{\dagger }{X}{}_{\mathbf{k}-\mathbf{q}}^{{{P}_{h'''}}} \right\rangle  \right\rangle }_{z}} \right],
\end{align}
\label{PopNCA}
\begin{align}
\Pi _{\mathbf{q}}^{{{N}_{v}}{{N}_{v}}}\left( \omega  \right)={{ F_{V}^{{{N}_{v}}} F_{V}^{{{N}_{v'}}} }}\sum\limits_{{\mathbf{p}},{N}_{v}}{{{\left| M_{\mathbf{p},-\mathbf{q}}^{bd} \right|}^{2}}{\Gamma _{{{N}_{v}}}^{A}}\overset{\dagger }{\Gamma}{}_{{{N}_{v'}}}^{A}} \times  \\ \nonumber
 \int{\int{\frac{d{{z}_{1}}d{{z}_{2}}}{{{\pi }^{2}}}{{f}_{V}} \left( \omega ,{{z}_{1}},{{z}_{2}} \right)}}
 \left[ \sum\limits_{\begin{smallmatrix} 
 {{P}_{h}},{{P}_{h'}} \\ 
 {{P}_{h''}},{{P}_{h'''}} 
\end{smallmatrix}}{\Gamma _{{{P}_{h}}{{P}_{h'}}}^{bd}}\overset{\dagger }{\Gamma}{}_{{{P}_{h''}}{{P}_{h'''}}}^{bd}\,\operatorname{Im}{{\left\langle \left\langle  X_{\mathbf{p}-\mathbf{q}}^{{{P}_{h''}}} | \overset{\dagger }{X}{}_{\mathbf{p}-\mathbf{q}}^{{P}_{h}} \right\rangle  \right\rangle }_{{{z}_{1}} }}\,\operatorname{Im}{{\left\langle \left\langle  X_{\mathbf{p}}^{{{P}_{h'}}} | \overset{\dagger }{X}{}_{\mathbf{p}}^{{{P}_{h'''}}} \right\rangle  \right\rangle }_{{{z}_{2}} }} \right],
\end{align}
\end{subequations}
\end{widetext}
with \[ {{f}_{H}}=\frac{1-{{n}_{F}}\left( z \right)+{{n}_{B}}\left( \Omega  \right)}{\omega -z-\Omega }\  \text{and} \ {{f}_{V}}=\frac{{{n}_{F}}\left( {{z}_{1}} \right)-{{n}_{F}}\left( {{z}_{2}} \right)}{\omega +{{z}_{1}}-{{z}_{2}}},\]
where ${{n}_{F}}\left( z \right)$ and ${{n}_{B}}\left( \Omega \right)$ are the Fermi- and Bose-distribution functions, respectively. Throughout the paper the frequency argument $\omega$ of the Green's functions is to be understood as $\omega +i0^+$. The diagonal EPI is omitted in the operator $\hat{\Pi}_{\mathbf{q}} \left( \omega  \right)$, it turned out that renormalization cased by this contribution are at least an order of magnitude smaller then the off-diagonal ones. 

To determine the electron and phonon Green's functions given by Eqs.~\ref{ElGrX} and \ref{PhGrX}, respectively, we need to solve the system of Eqs.~\ref{DysonEl}-\ref{PopNCA}. After the exact diagonalization of the Hamiltonian $H_c$ from Eq.~\ref{cTransfH} for the CuO$_4$ unit cell and transformations to the forms of Eqs.~\ref{X_Hc},\ref{X_Hcc}, we carry out the simulations. For self-consistency cycles we use the 200$\times$200 $\mathbf{k}$ points in the Brillouin zone in the generalized mean-field approximation and then proceed in the full scheme of Eqs.~\ref{DysonEl}-\ref{PopNCA} using the 20$\times$20 $\mathbf{k}$-points in the Brillouin zone. To consider the electron band structure of polarons and the phonon band dispersion we find the poles of the Fourier transformed Green's functions $G_{\mathbf{k},\sigma }^{\beta \beta '}\left( \omega  \right)$ and ${{D}_{\mathbf{q}}}\left( \omega  \right)$. We also define the corresponding densities of states, for the phonon excitations it reads ${{N}_{ph}}\left( \omega  \right) = \frac{1}{N}\sum\limits_{\mathbf{q}}{{B}_{\mathbf{q}}\left( \omega \right)}$, where ${B}_{\mathbf{q}}\left( \omega \right)$ is the spectral function of phonons, which is defined as
${{B}_{\mathbf{q}}}\left( \omega  \right)=-\frac{1}{\pi }\operatorname{Im}{{D}_{\mathbf{q}}}\left( \omega +i\delta  \right)$. In a similar way one-electron density of states at the Fermi level is given by ${{N}}\left( \omega = E_f  \right) = \frac{1}{N}\sum\limits_{\mathbf{k}}{{A}_{\mathbf{k},\sigma}\left( \omega \right)}$, where ${{A}_{\mathbf{k},\sigma}}\left( \omega  \right)=-\frac{1}{\pi }\sum\limits_{\beta}{\operatorname{Im}{{G}_{\mathbf{k},\sigma}^{\beta,\beta}}\left( E_f +i\delta  \right)}$.

\section{Computational details \label{sec_Computer}}
In contrast to Lanczos method we exactly determine all eigenstates of the cluster Hamiltonian $H_c$ at the stage of its diagonalization. The GTB approach allows one to discard excited states in a controlled manner, which does not lead to significant changes of the spectrum. The preliminary computations show that transitions related with highly excited states, as well as transitions between sectors of the Hilbert space with $n_h=2$ and $n_h=3$, do not affect the dispersion of charge carriers near the Fermi level. To obtain the correct dispersion in a system with EPI, we consider how the energy of local eigenstates depends on the number of phonons taken into account. It turns out that the best criterion is the changes in the energy of the ground and the first excited states in sectors with $n_h=0,1,2$. Therefore, we increase the number of phonons and determine the value of $N_{max}$ so that calculation with $N_{max}+1$ varies these energies less than $ 1 \% $. The typical values for achieving such convergence range from $N_{max}=40$ to $N_{max}=400$. The number of excited states under consideration depends on the model parameters and is determined empirically, but we still neglect the transitions that form the high-energy or far from the Fermi level part of the band structure and do not affect the dispersion and redistribution of the spectral weight near the Fermi level. Note the total spectral weight of each subband for a given spin projection in the presented band structure calculations differs from unity by no more than $0.005$.

\section{Polaron band structure \label{sec_BandStrc}}
\begin{figure}
\includegraphics[width=\linewidth]{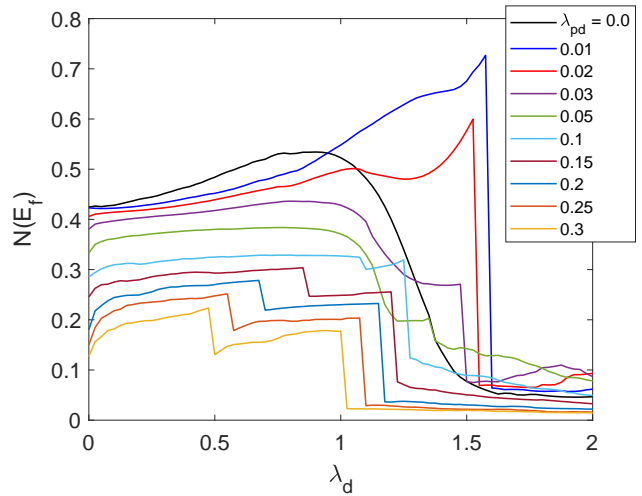}
\caption{\label{fig:f1}
The density of states at the Fermi level $N\left({E_{f}}\right)$ for different values of diagonal $\lambda_d$ and off-diagonal $\lambda_{pd}$ parameters of EPI.}
\end{figure}

\begin{figure*}
\center
\includegraphics[width=\linewidth]{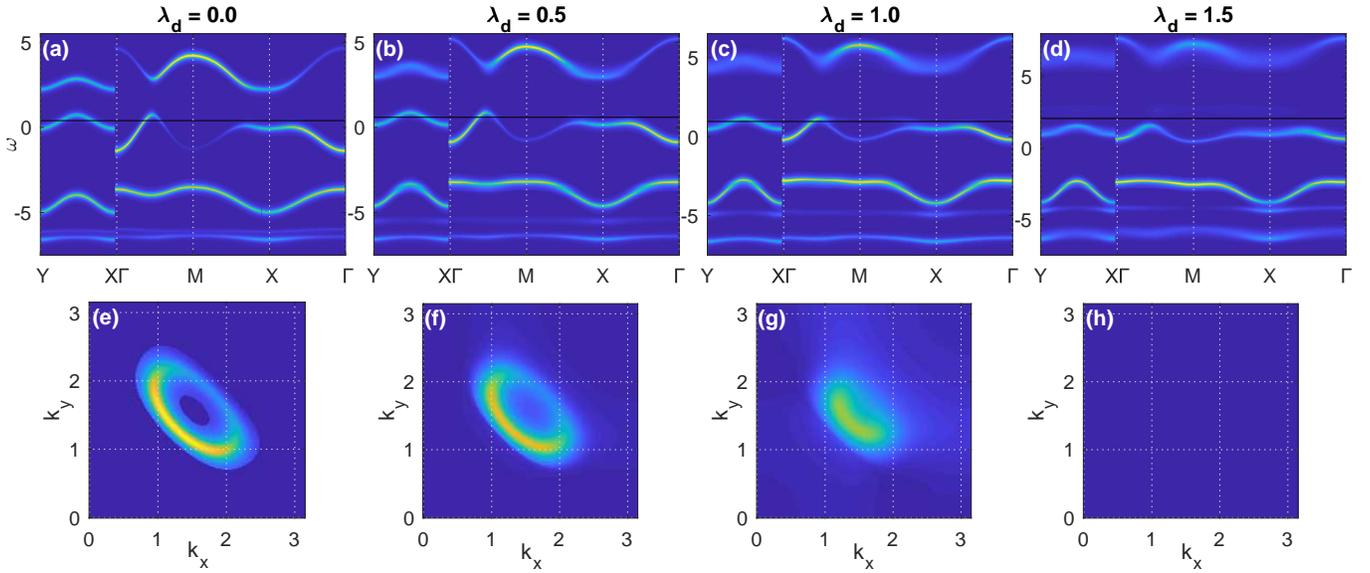}
\caption{\label{fig:f2}
The smooth evolution of the band structure (a,b,c,d) and gradual changing of the spectral weight mapping in k-space at the Fermi level (e,f,g,h) for purely diagonal electron-phonon interaction ($\lambda_{pd}=0$). The chemical potential is marked by a black line. Here and throughout the figures, the energy values are given in eV and the concentrations $x$ of doped holes is $0.07$.}
\end{figure*}

\begin{figure*}
\center
\includegraphics[width=\linewidth]{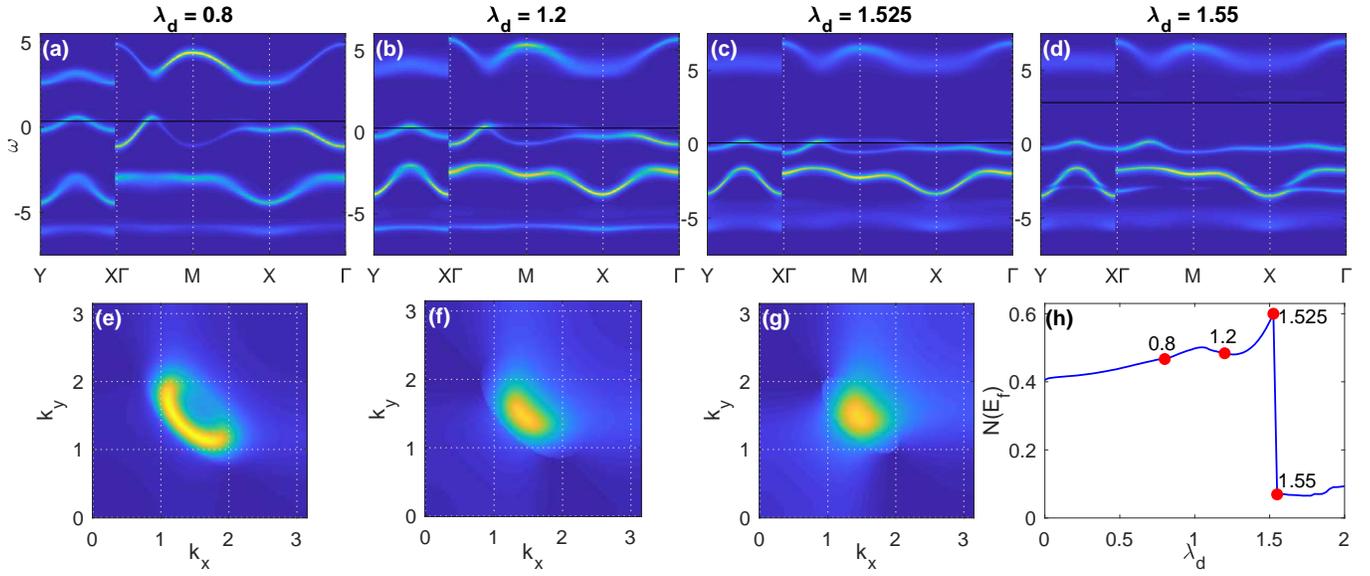}
\caption{\label{fig:f3}
The typical transformation of the band structure (a,b,c,d) and the spectral weight mapping in k-space at the Fermi level (e,f,g) for parameter range $0< {{\lambda }_{d}}\le 2$ and $ 0 < {{\lambda }_{pd}} < {\lambda }_{pd}^c $, which corresponds to smooth changing of the local polaron properties, here $ \lambda_{pd} = 0.02 $. The increasing of diagonal contribution at fixed off-diagonal one leads to the quantum phase transition and transformation of the Fermi surface from electron hole pockets centered at $\left( {\pi }/{2};{\pi }/{2} \right)$ point of the Brillouin zone to Fermi arcs centered at $\left( \pi ;\pi  \right)$ point. The density of states at the Fermi level indicates sharp transition from delocalized to localized state (h).}
\end{figure*}

\begin{figure}
\center
\includegraphics[width=\linewidth]{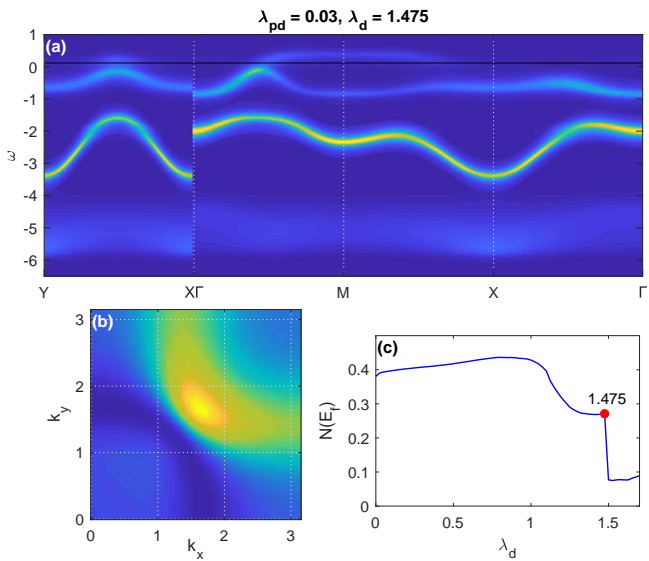}
\caption{\label{fig:f4}
The band structure (a), spectral weight mapping in k-space (b) and density of states at the Fermi level (c) for the EPI parameters corresponding to formation of blurred and elongated Fermi arcs (weak pseudogap limit).}
\end{figure}

\begin{figure*}
\center
\includegraphics[width=\linewidth]{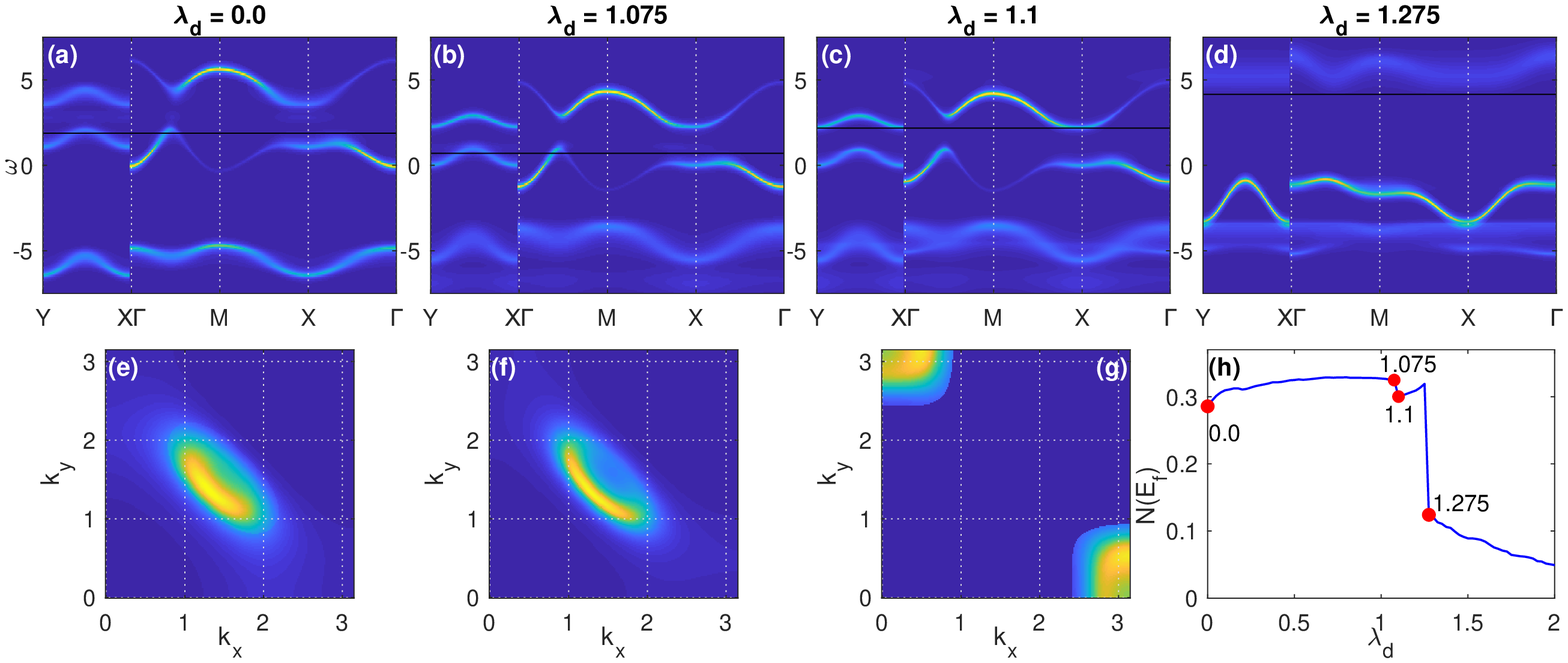}
\caption{\label{fig:f5}
The typical transformation of the band structure (a,b,c,d) and spectral weight mapping in k-space at the Fermi level (e,f,g) for the EPI parameter range $0 < {{\lambda }_{d}}\le 2$ and $\lambda_{pd} > \lambda_{pd}^c $, where sharp transitions in the local and lattice polaron properties are observed, here $\lambda_{pd}=0.1$. The corresponding density of states at the Fermi level (h) undergoes two sharp drops. The first of them corresponds to the quantum phase transition related with the transformation of Fermi surface from a hole to an electron-like one. The second drop of the $N\left({E_{f}}\right)$ is a sign of the metal-insulator transition.}
\end{figure*}

The phase diagram of the copper oxides reveals evolution from an undoped antiferromagnetic insulator to an almost conventional Fermi liquid for the overdoped system. Contemporaneously strong effects of electron correlations and electron-phonon couplings are most significant in underdoped cuprates, so we consider the compound La$_{2-x}$Sr$_x$CuO$_4$ with a low level of hole doping, $x=0.07$. Figures~\ref{fig:f1},~\ref{fig:f2},~\ref{fig:f3},~\ref{fig:f4},~\ref{fig:f5} show the calculated density of states at Fermi level, bands along the directions in the Brillouin zone with the chemical potential marked by black line, and spectral weight mapping in k-space at Fermi level.

First of all we analyze the band structure of the system without electron-phonon interaction. We have a charge transfer insulator for the half-filling case. For a hole doped compound with concentration of doping $x=0.07$, we find the chemical potential lying in the upper part of the valence band (Figs.~\ref{fig:f2}~(a) and \ref{fig:f2}~(e)). So, it is a metal without electron-phonon coupling. Instead of wide tight-binding bands we observe the more narrow Hubbard subbands with inhomogeneous in k-space spectral weight redistribution between them. In contrast to paramagnetic dispersion with maximum of the valence band at the $M\left( \pi ; \pi  \right)$ point of the Brillouin zone, the maximum of the valence band of the underdoped compound with short-range antiferromagnetic order lies near the point $\left( {\pi }/{2} ; {\pi }/{2} \right)$. Like other transition metal oxides the underdoped cuprates are characterized by strong interaction of charge carriers with spin fluctuations. This interaction induces inhomogeneous in k-space spectral weight redistribution between Hubbard subbands and leads to the formation of the shadow band in the short range antiferromagnetic spin liquid state, where instead of umklapp process with momenta of order $\left( \pi ;\pi  \right)$, dynamic decay occurs with a finite lifetime. As a result a spectral weight mapping in k-space at the Fermi level demonstrates small hole pockets centered at $\left( {\pi }/{2} ;{\pi }/{2} \right)$ point of the Brillouin zone and characterized by a larger spectral weight on the outer side of the Fermi contour (Figs.~\ref{fig:f2}~(e)).

The electron-phonon interaction leads to the hybridization between Hubbard fermion subbands and Frank-Condon resonances splitting the former into a number of the Hubbard polaron subbands~\cite{PhysRevB.92.155143}. The process is accompanied by significant transformation of the bands and Fermi surface (Figs.~\ref{fig:f2}-\ref{fig:f5}). Besides the spectral weight transfer to the high-energy multiphonon states and the related loss of the coherence of the bands, in a certain region of the EPI parameters we observe the formation of the flat band close to the Fermi level and the subsequent pinning of the chemical potential in it (Figs.~\ref{fig:f3}~(b),~\ref{fig:f3}~(c), and~\ref{fig:f4}~(a)). One more drastic effect is the transition from metallic to dielectric state in the limit of strong electron-phonon interaction, which is clearly seen from the density of state at the Fermi level (Fig.~\ref{fig:f1}). As the EPI increases, the bands narrow and the chemical potential is pushed into the gap (Figs.~\ref{fig:f2}~(d),~\ref{fig:f3}~(d), and ~\ref{fig:f5}~(d)), indicating the localization of the polaron charge carriers.

\begin{figure*}
\center
\includegraphics[width=\linewidth]{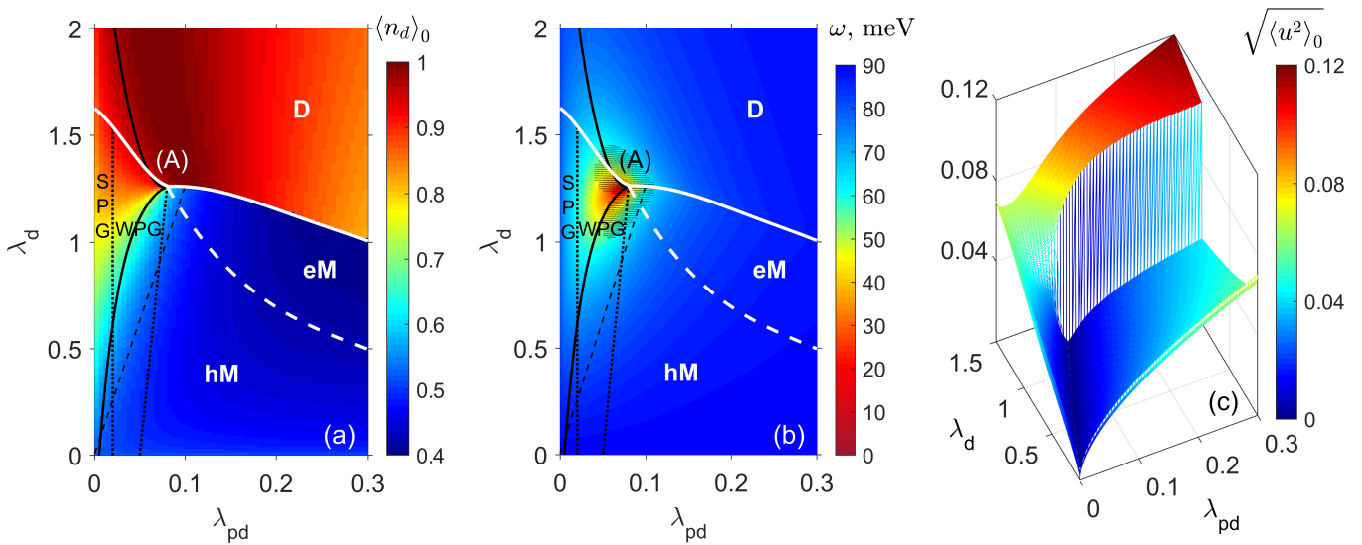}
\caption{\label{fig:f6}
The phase diagram of the polaron system in the plane of diagonal and off-diagonal EPI parameters mapped to (a) local polaron property $\left\langle {{n}_{d}} \right\rangle$ and (b) the shifted frequency of the phonon excitation with the highest intensity for the point X of the Brillouin zone, (c) the square root of the average square of oxygen displacement $\sqrt {{{\left\langle {{u^2}} \right\rangle }_0}} $ in the unit of the lattice constant $a$. Here eM (hM) is the metal state with electron (hole) charge carries, D is the dielectric state, (A) is the critical point. White solid curve indicates the transition between delocalized and localized polaronic states. Black solid curve marks the frequency with maximal shift from the bare phonon line at each fixed parameter $\lambda_d$. Black dashed line is a ``compensation parameter line''. (More explanations are given in the text.)}
\end{figure*}

The competition of diagonal $\lambda_d$ and off-diagonal $\lambda_{pd}$ electron-phonon contributions defines the character of the polaron band structure transformations. To demonstrate it we consider a wide range of parameters and define on Fig.~\ref{fig:f6} the diagram of the polaron properties in the plane of the values $\lambda_{d}$ and $\lambda_{pd}$, varying them as $0\le {{\lambda }_{d}}\le 2$ and $0\le {{\lambda }_{pd}}\le 0.3$.  First of all we analyze one-electron density of polaronic states on the Fig.~\ref{fig:f1} and reveal that increasing of EPI strength results in the smooth changes of the density of states at the Fermi level if purely diagonal contribution is taken into account (curve for $\lambda_{pd}=0$). The sharp changes with one or two collapses occur if both diagonal and off-diagonal interactions are involved. The related types of the band structure and Fermi surface evolution are presented below.

The first type of evolution (i) is shown on the Fig.~\ref{fig:f2} and corresponds to purely diagonal electron-phonon contribution, $\lambda_{pd}=0$. With increasing of EPI strength $\lambda_d$ we observe the uniform decreasing of the spectral weight on both sides of the hole pockets up to the metal-insulator transition where the Fermi surface disappears (Fig.~\ref{fig:f2}~(h)). At sufficiently large value of $\lambda_d$, the Fermi surface looks like an arch (Fig.~\ref{fig:f2}~(c)), but indeed this is a hole pocket with inhomogeneously distributed spectral weight.

The second type (ii) of the band structure and Fermi surface evolution takes place for $0 < {{\lambda }_{pd}} < \lambda_{pd}^c $, where critical value is ${{\lambda }_{pd}^{c}} \approx 0.05$. The evolution is characterized by transformation of small hole pockets to Fermi arcs before the metal-insulator transition (Fig.~\ref{fig:f3}). Let us consider the evolution in details by increasing the value of ${{\lambda }_{d}}$ at any fixed value of ${{\lambda }_{pd}}$ from the above limits. At weak electron-phonon interaction, we again observe small hole pockets with inhomogeneously distributed spectral weight (Fig.~\ref{fig:f3}(a),(e)). However, increasing of the diagonal contribution ${{\lambda }_{d}}$ results in the formation of the flat band close to the Fermi level at intermediate and strong electron-phonon coupling strength. It is clearly seen along $\Gamma \left( 0;0 \right)$ - M $\left( {\pi }/{2} ;{\pi }/{2} \right)$ - X$ \left( {\pi } ; 0 \right)$ directions of the Brillouin zone in the Figs.~\ref{fig:f3}~(b),(f) and ~\ref{fig:f3}~(c),(g). This effect enhances the spectral weight redistribution between the inner and outer sides of the hole pockets while the chemical potential is in the valence band, since a part of the spectral weight is transferred  to the flat band. However we find the finite spectral weight at the inner parts of the Fermi pockets until the chemical potential enters the flat band. From this point on, the topology of the Fermi surface changes and instead of small hole pockets centered at the $\left( {\pi }/{2} ;{\pi }/{2} \right)$ point of the Brillouin zone (Figs.~\ref{fig:f3}~(e)), we observe the Fermi arcs centered at $\left( \pi ;\pi  \right)$ point and having a maximum spectral weight at their center (Figs.~\ref{fig:f3}~(f) and ~\ref{fig:f3}~(g)).

The reconstruction of the band near the $\left( \pi ;\pi  \right)$ point of the Brillouin zone seems to be related with destruction of the short range antiferromagnetic order and subsequent hybridization between fragments of the restored paramagnetic band and Frank-Condon resonances. The flat band itself is a characteristic feature of polaron systems~\cite{PhysRevB.73.205122} and it indicates the formation of heavy polarons. Indeed, we find the flat bands at intermediate or strong electron-phonon coupling limits (Figs.~\ref{fig:f3}~(b),(c) and \ref{fig:f4}~(a)). At the same time, it results from the diagonal electron-phonon interaction, when off-diagonal ones are rather small. It is naturally to assume that the increase in the diagonal contribution related to the charge density fluctuation on copper leads to the damping of the short range antiferromagnetic order. For sure, the sharp drop of the electronic density of states at the Fermi level (Figs.~\ref{fig:f4}~(c)) will result in the strong decreasing of the magnetic susceptibility. More detailed analysis is planned in the future work.
Note that the similar effect has been recently observed for the temperature and doping evolution of the band structure in the Hubbard model~\cite{PhysRevB.90.245104,PhysRevB2020}. When the chemical potential enters into the flat band and the quantum phase transition is observed, the density of states raises sharply (Fig.~\ref{fig:f1}, curves for $\lambda_{pd} = 0.01,\ 0.02$ and Fig.~\ref{fig:f3}~(h)) or becomes approximately constant (Figs.~\ref{fig:f1}, curves for $\lambda_{pd} = 0.03,\ 0.05$ and Figs.~\ref{fig:f4}~(c)) up to the metal-insulator transition. The profiles of the Fermi arcs in these regimes differ significantly. In the first case, we find short and clearly defined arcs (Figs.~\ref{fig:f3}~(f),(g)), and in the second, arcs are blurred and elongated to the boundaries of the Brillouin zone (Fig.~\ref{fig:f4}~(b)). In accordance with Ref.~\cite{PhysRevB2020}, we refer these states to the strong or weak pseudogap limit, respectively. A further increasing of the EPI leads to the sharp localization of the polaronic charge carries, as a result the density of states at the Fermi level drops to zero and the Fermi surface disappears (Figs.~\ref{fig:f3}~(d),(h) and~\ref{fig:f4}~(c)).

The third type (iii) of the band structure and Fermi surface evolution is shown on the Figure~\ref{fig:f5} and takes place approximately for $\lambda_{pd} > \lambda_{pd}^c $. Similar to the case (i) the uniform decreasing of the spectral weight on the Fermi surface is observed with increasing of the diagonal contribution $\lambda_{d}$ in the limits from the weak to intermediate electron-phonon coupling (Figs.~\ref{fig:f5}~(a),(e) and Figs.~\ref{fig:f5}~(b),(f)). The main transformations of the band structure is observed far from the Fermi surface for the lower bands, which become incoherent even at weak EPI. As the value of $\lambda_{d}$ increases, the electron-phonon interaction pushes the chemical potential out of the valence band through the gap into the conductivity band (Figs.~\ref{fig:f5}~(b),(c)), so the Fermi surface changes its topology and we find the electron pockets at the X$\left( \pi ;0  \right)$ and Y$\left( 0 ;\pi  \right)$ point of the Brillouin zone (Fig.~\ref{fig:f5}~(g)). The process is accompanied by sharp drop of the density of states at the Fermi level to its intermediate value (Fig.~\ref{fig:f5}~(h) and Fig.~\ref{fig:f1} for curves with $\lambda_{pd} \ge 0.1$). Thus, at sufficiently large contribution of off-diagonal EPI, the diagram of polaron properties includes the parameter regions of metal states with different types of charge carries. Continuing to increase the parameter of diagonal interaction, we reach the metal-insulator transition and observe the second sharp drop in the density of states (Fig.~\ref{fig:f5}~(h)). Comparing the Figs.~\ref{fig:f3} and~\ref{fig:f5} we conclude that for ${{\lambda }_{pd}}\le {\lambda }_{pd}^c$, with an increase of $\lambda_d$, the upper band first becomes incoherent, and then the Fermi level enters into the gap. While for the ${{\lambda }_{pd}} > {\lambda }_{pd}^c$, everything happens the other way around. The competition of the diagonal and off-diagonal electron-phonon contributions preserves the coherence of the upper band up to the metal insulator transition and therefore provides the appearance of the electron metal state.

The results of this section are summarized on the phase diagram in the plane of EPI parameters $\lambda_d$ and $\lambda_{pd}$ (Fig.~\ref{fig:f6}~(a)). Here white solid line marks the metal-insulator transition in accordance with the computations of the density of states at the Fermi level (Fig.~\ref{fig:f1}). White dashed line divides metallic states with different kind of conductivity. The parameter space corresponding to the second (ii) and third (iii) types of the band structure and Fermi surface evolution are separated by the slanted black dotted line ending at the point (A). The regions of the strong and weak pseudogap states are divided by the vertical black dotted line. 

Since polaron properties are strongly depends on local polaron structure, described by Eq.~\ref{pstates}, we also display the average number of holes on the d-orbital for the single-hole ground state ${\left\langle {{n}_{d}} \right\rangle}_{0}$. Being a kind of polaron localization measure this local function characterizes the electron-phonon coupling strength. At zero electron-phonon interaction the distribution of the holes among copper and oxygen orbitals corresponds to $\lambda_d=\lambda_{pd}=0$ point of the phase diagram in the Fig.~\ref{fig:f6}~(a). The stronger the EPI becomes, the more charge carriers are in the copper orbitals. There are two regimes of the value ${\left\langle {{n}_{d}} \right\rangle}_{0}$ changes~\cite{Mak2016}. For $\lambda_{pd} < \lambda_{pd}^c$, we find smooth behavior, although the changes are rather fast for the intermediate and strong electron-phonon coupling near and at the metal-insulator transition. For $\lambda_{pd} > \lambda_{pd}^c$, sharp and discontinuous changes accompany the transition from the delocalized to the localized state. The color mapping of the value ${\left\langle {{n}_{d}} \right\rangle}_{0}$ in the Fig.~\ref{fig:f6}~(a) clearly demonstrates the critical point (A) where abrupt changes of the local polaron property begin to develop. Note that other local polaron properties, such as the number of phonons $N_{max}$ involved in the formation of polaronic state, the maximum of hole distribution among states with different phonon number, and the square root of average square of oxygen displacement $\sqrt {{{\left\langle {{u^2}} \right\rangle }_0}} $ at given strength of EPI, have similar behavior. As an example, we display the value $\sqrt {{{\left\langle {{u^2}} \right\rangle }_0}} $ in the Fig.~\ref{fig:f6}~(c). Again, for a purely diagonal EPI contribution, we find a smooth behavior that persists with a small fixed off-diagonal EPI contribution up to the point (A) where sharp changes occur. Summarizing, we observe rapid but smooth increasing of the local averages to the left of the critical point (A), where diagonal contribution dominates, and their discontinuity to the right of this point, where off-diagonal interaction is prevalent. At the same time the transformation of the polaronic band structure is smooth only for diagonal EPI. So the first type (i) of the band structure and Fermi surface evolution takes place when both local and lattice polaron properties change smoothly. The second type of evolution (ii) is realized when lattice polaron properties change sharply but local one change smoothly. The third type (iii) of evolution corresponds to the sharp changes in the local and lattice properties simultaneously.

\section{Phonon spectral function \label{sec_PhSF}}
\begin{figure}
\center
\includegraphics[width=\linewidth]{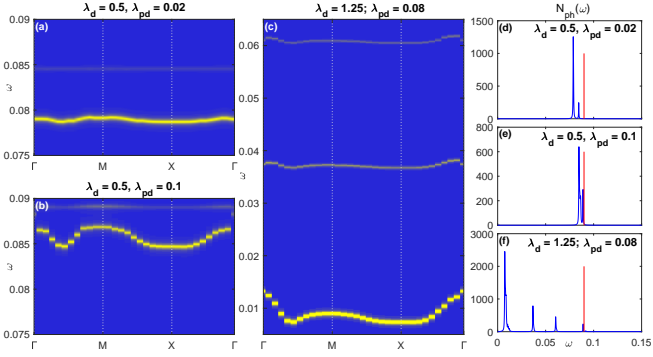}
\caption{\label{fig:f7}
Dispersion of the phonon frequency (in eV) for some characteristic points of the phase diagram, EPI parameters correspond to (a) ``compensation line'', (b) weak coupling regime, and (c) region near the critical point (A). (d,e,f) The corresponding density of states of the phonon excitations of polaron. Red line indicates the bare phonon frequency $\omega_{0}$.}
\end{figure}

Here we consider the phonon excitations of polarons. The coupling of electrons and phonons results in the phonon band dispersion, the shift of the bare phonon line away from its resonant value $\omega_0$, and the emergence of the new states in the phonon spectrum. If electron-phonon interaction is absent in the total Hamiltonian then spectral function ${{B}_{\mathbf{q}}}\left( \omega  \right)$ describes dispersionless excitation with the bare phonon frequency $\omega_0$  (Eq.~\ref{FreePhononGF}). Weakly dispersive spectrum $\Omega \left( \mathbf{k} \right)$ as in the Fig.~\ref{fig:f7}~(a) is observed along and near the black dashed line in the phase diagram on Fig.~\ref{fig:f6}. It turns out there is significant compensation of the diagonal and off-diagonal EPI contributions there. Indeed, the structure of the local polaronic states corresponding to this line is characterized by a small number of the phonons $N_{max}$ and a distribution of the density of charge carriers with a maximum lying on the states with $n_{ph}=0$ or $n_{ph}=1$ (Eq.~\ref{pstates}). As a result, the square root of the average square of oxygen displacement $\sqrt {{{\left\langle {{u^2}} \right\rangle }_0}} $ takes its minimum values along the ``compensation line'' that is seen from the Fig.~\ref{fig:f6}~(c). 

Throughout the phase diagram the momentum dependent spectrum demonstrates the largest softening of the phonon frequency at the X and Y points at the edges of the Brillouin zone (Fig.~\ref{fig:f7}~(b) and (c)). Such frequency behavior is in a qualitative agreement with experimental data for the breathing mode in copper oxide systems \cite{doi:10.1002/pssb.200404951,PhysRevB.60.R15039,Reznik2010}. In our model the dispersion $\omega \left( {\mathbf{q}} \right)$ results from off-diagonal electron-phonon interaction and is defined by the features of the matrix element $M_{\mathbf{q},\mathbf{k}-\mathbf{q}}^{pd}$ and polaron band structure in the system with short range antiferromagnetic order (Eq.~\ref{PopNCA}). It has been shown earlier that non-local electron-phonon interaction effects lead to the anomalies of the bond-stretching mode in cuprates \cite{PhysRevB.64.054516}. Nevertheless, a detailed analysis of the peculiarities of the phonon spectrum requires a more thorough approach here, taking into account, in addition to electron-phonon scattering, inelastic scattering on spin and charge excitations of the system. This problem, as well as the effects of doping and temperature, will be the subject of further research. 

A non-zero electron-phonon interaction induces the transitions between ground and excited polaronic states, giving rise to the off-resonant part of the phonon spectral function. Low-energy excited states formed by a polaron and additional phonon quanta can be classified as bound, unbound, and antibound in accordance with the energy value which separates them from the ground state energy. Thus, besides the phonon excitations with the resonant value $\omega_0$, we find the states below and above the one-phonon continuum in the structure of the phonon spectral function. For the entire parameter range under study, the transitions between ground and antibound polaronic states with energies higher than $ \omega_0 $ have a vanishingly small spectral weight. The main spectral weight belongs transitions between ground and bound or unbound states. Usually we observe several excitations with comparable spectral weights (Figs.~\ref{fig:f7}~(a),(b) and~(c)) but it is difficult to resolve additional excitations from the main peak with the highest intensity (Figs.~\ref{fig:f7}~(d) and (e)).

The situation drastically changes at intermediate or strong electron-phonon coupling strength near the critical point (A) where local polaron properties and polaron structure start to change fast or even abruptly. Then, below the phonon threshold (i.e., the minimum inelastic scattering energy), the phonon spectral function demonstrates several, namely 1,2 or 3 additional phonon excitations which are well separated from each other and from the main peak with the highest intensity (Fig.~\ref{fig:f7}~(c)). The parameter range, where these novel states emerge in the spectrum, is shaded on the Fig.~\ref{fig:f6}~(b). Previously the comprehensive study of the novel states in the phonon spectral function of the Holstein polaron has been reported for the weak, intermediate, and strong coupling regimes of both adiabatic and non-adiabatic systems \cite{Loos2006,PhysRevB.73.214304,PhysRevB.82.104304,PhysRevB.91.165127}. However, in these papers only the diagonal electron-phonon contribution has been taken into account, while the emergence of the novel states revealed here results from the critical behavior of the system due to the competition of the diagonal and off-diagonal electron-phonon interactions. On the other hand, the obtained feature of the phonon spectral function reflects the structure of low-energy excited polaronic states in the crossover regime. In polaron models with other types of short-range electron-phonon interaction, also up to 3 excited states have been found below the one-phonon continuum from the onset of the strong coupling regime~\cite{PhysRevB.66.020301,PhysRevB.69.064302,PhysRevB.99.134308}. So this appears to be a generic characteristic of such systems.

The frequencies of phonon excitations with the most intensity at given diagonal and off-diagonal EPI parameters are presented in the Fig.~\ref{fig:f6}~(b) for one chosen point of the Brillouin zone. Even though the largest contribution to the energy shift of the bare phonon frequency comes from the diagonal electron-phonon interaction, the maximal ``renormalizations'' at each fixed parameter $\lambda_d$ are defined by the competition of diagonal and off-diagonal ones. It is shown by black solid curve on the Figs.~\ref{fig:f6}~(a) and~(b) which is turned out to be located in the region of smooth evolution of local properties. Figure~\ref{fig:f6}~(b) also demonstrates that the closer EPI parameters are to the critical point, the stronger phonon frequency shift is observed. Note that the value of the frequency shift near the critical point (A) depends on dimension of the Hilbert space in our calculations. Therefore, we do not confidently discuss the nature of excitations with frequency tending to zero at certain momenta.

\section{Discussion and Conclusions\label{sec_Conclusion}}
In this paper we study the multiband system of strongly correlated electrons coupled with one optical phonon mode. The diagram of polaron properties is defined in the many-body scenario for a wide range of electron-phonon parameters from weak to strong coupling limit at adiabatic ratio $t \gg \omega_0$. We demonstrate how competition between the diagonal (local) and off-diagonal (transitive) contributions of electron-phonon interaction defines the critical behavior of the system. The smooth crossover from delocalized to localized state and gradual evolution of the polaron properties are observed only for the purely diagonal electron-phonon interaction. Then increasing of EPI parameter strength $\lambda_d$ leads to the smooth and continuous changes in the density of states at the Fermi level and Fermi surface. Also, local polaron properties, for example, average number of holes on the d-orbital for the single-hole ground state ${\left\langle {{n}_{d}} \right\rangle}_0$ and square root of the average square of oxygen displacement change gradually and continuously.

The interference of the diagonal and off-diagonal electron-phonon contributions gives rise to the appearance of the critical point on the polaron phase diagram A$\left( \lambda_{d}^c,\lambda_{pd}^c \right)$. With increasing strength of the diagonal coupling for off-diagonal values $ 0 < {{\lambda }_{pd}} < \lambda _{pd}^{c} $, we find smooth behavior of the local features, but abrupt changes of the density of states. Before transition from a delocalized state to a localized one, the system undergoes quantum phase transition due to the transformation of the Fermi surface from small hole pockets centered at $\left( {\pi }/{2};{\pi }/{2} \right)$ point of the Brillouin zone to Fermi arcs centered at $\left( \pi ;\pi  \right)$ point. It should be noted that Fermi arcs form for the part of the phase diagram, where diagonal contribution dominates, namely, on the left from the ``compensation line'', indicated as the black dashed line on the Figs.~6a and~6b. However, the off-diagonal contribution, which depends not only on transferred momentum $\mathbf{q}$, but also on initial momentum $\mathbf{k}$, is indispensable ingredient of that transformation. Moreover, strength of the off-diagonal contribution determines the shape of Fermi arcs which reflects the strong or weak pseudogap limit of the system. To the right of the critical point (A), for  $\lambda _{pd}^{c} < {{\lambda }_{pd}}$, the off-diagonal contribution dominates. At this part of the phase diagram, both local and lattice polaron properties are characterized by the sharp changes when the metal-insulator transition occurs. Before transition to a localized state the increasing of the diagonal contribution leads to the quantum phase transition caused by the transformation from hole to electron like Fermi surface. This band structure transformation is accompanied by the sharp drop of the density of states at Fermi level to its intermediate value.

The properties of the phonon spectral function of the polaron are also influenced by the competition of the diagonal and off-diagonal electron-phonon interactions. We find that the largest contribution to the energy shift of the bare phonon frequency comes from the diagonal electron-phonon interaction, while the off-diagonal one mainly causes the phonon dispersion. The most pronounced effects are observed around the critical point on the phase diagram on the Fig.~\ref{fig:f6}, where maximal phonon frequency shifts and emergence of the novel states below the one-phonon continuum occur.

The found phase diagram has revealed that both diagonal and the off-diagonal electron-phonon contributions can critically impact the behavior of the system with strong electron correlations. We have shown how electron-phonon interaction leads to the formation of the pseudogap state. The corresponding transformation of the band structure is controlled by the dominating diagonal contribution, but the off-diagonal one switches the strong or weak pseudogap regimes. When off-diagonal electron-phonon interaction is prevalent, the phase diagram includes the region where quantum phase transition from the hole type of metal to the electron one occurs. These transformations, i.e. the pseudogap formation or the change of conductivity type, are separated by the non-analyticity point on the phase diagram and accompanied also by the features of the phonon spectral function near this point.

\begin{acknowledgments}
We acknowledge the stimulating discussions with V.A. Gavrichkov, G. Seibold, and K.I. Kugel.
The reported study was funded by Russian Foundation for Basic Research, Government of Krasnoyarsk Territory and Krasnoyarsk Regional Fund of Science according to the research project ''Features of electron-phonon coupling in high-temperature superconductors with strong electronic correlations'' No. 18-42-240017 and project  "Electronic  correlation  effects  and multiorbital physics in iron-based materials and cuprates" number 19-42-240007".
\end{acknowledgments}

\bibliography{mybibfile}
\end{document}